\documentclass{memtensor}

\usepackage[utf8]{inputenc}
\usepackage{inconsolata}
\usepackage{url}
\usepackage{marvosym}

\makeatletter
\def\bstctlcite{\@ifnextchar[{\@bstctlcite}{\@bstctlcite[@auxout]}}
\def\@bstctlcite[#1]#2{\@bsphack
  \@for\@citeb:=#2\do{%
    \edef\@citeb{\expandafter\@firstofone\@citeb}%
    \if@filesw\immediate\write\csname #1\endcsname{\string\citation{\@citeb}}\fi}%
  \@esphack}
\makeatother

\title{RouteAlign: Aligning Task Requirements with Model Profiles for LLM Service Selection}

\author[1]{Jihao Zhao$^{*,}$}
\author[1]{Ding Chen$^{*,}$}
\author[2]{Yanfang Chen}
\author[1]{Bo Tang}
\author[1]{Feiyu Xiong}
\author[3]{Junchi Yan}
\author[1]{Zhiyu Li\textsuperscript{\Letter{},}}

\affiliation[1]{MemTensor (Shanghai) Technology Co., Ltd., Shanghai, China}
\affiliation[2]{Renmin University of China, Beijing, China}
\affiliation[3]{Shanghai Jiao Tong University, Shanghai, China}

\abstract{
LLM service selection requires performance evidence that reflects the demands of the current task. Aggregate scores obscure differences in capability and difficulty, while semantically related examples can involve different combinations of skills. We present RouteAlign, which uses task requirements to structure both the measurement of model performance and the selection of services. Task requirements describe capability demand and difficulty, enabling alignment with corresponding model performance measurements. Historical requirements determine how outcomes contribute to global profiles and local capability estimates; incoming requirements determine how those measurements inform the current decision. A shared scorer integrates the evidence, and a generation-cost preference controls selection among eligible services. Evaluation across diverse tasks and heterogeneous LLM services shows that RouteAlign extends the empirical quality-cost frontier toward higher task quality. Comparisons with open Jev-style decision backends show consistent quality gains, with lower generation cost than their strongest tested configuration. Ablations support the contribution of local performance evidence, while catalog-update experiments demonstrate effective reuse of the fitted scorer with refreshed model measurements.
}

\date{\today}
\correspondence{\email{lizy@memtensor.cn}}

\checkdata[Author Legend]{* Co-equal primary author, \Letter{} Corresponding author}
\checkdata[Code]{\url{https://github.com/MemTensor/RouteAlign}}
\checkdata[Keywords]{Large language models, service selection, task requirements, model profiles, quality of service.}

\begin{document}
\bstctlcite{RouteAlign:BSTcontrol}
\maketitle

\section{Introduction}
\label{sec:introduction}

Applications increasingly draw on large language models (LLMs) as services for tasks that span knowledge, reasoning, programming, and interaction with external systems. The breadth of foundation-model applications~\citep{bommasani2021foundation} makes the choice of service consequential: the same candidate may be suitable for routine requests yet unreliable on a demanding combination of skills. Service-oriented computing provides the architectural basis for invoking interchangeable functionality~\citep{papazoglou2007service}, and classical service selection compares eligible alternatives under quality-of-service (QoS) constraints~\citep{zeng2004qos}. For LLMs, however, task quality must be estimated for the particular request before the selected service executes it. An interface establishes how a model can be invoked; effective selection additionally requires evidence of what that model can accomplish under the current task demands.

\begin{figure}[t]
 \centering
 \includegraphics[width=0.47\columnwidth]{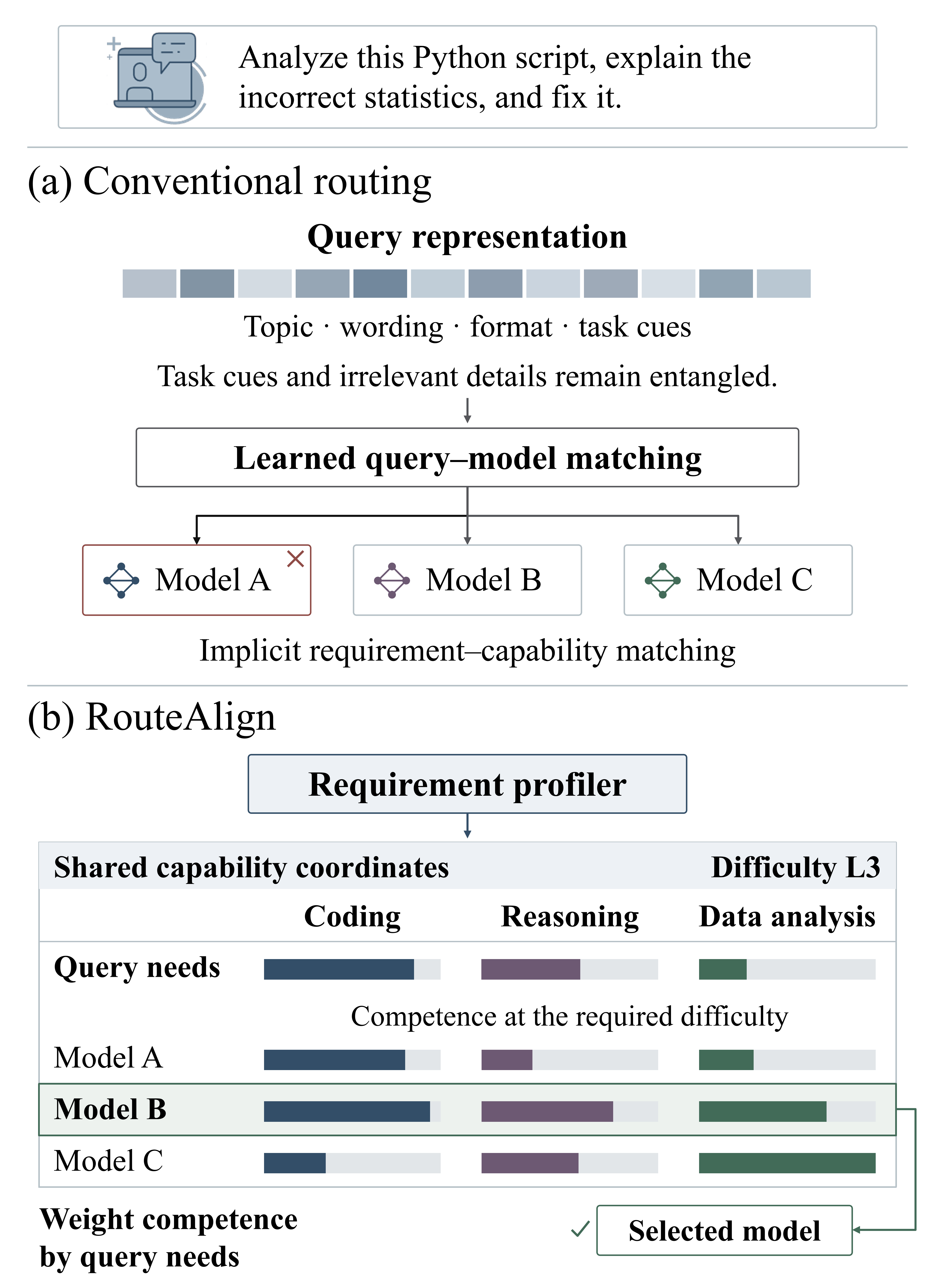}
 \caption{Conceptual comparison of query-model matching and requirement-profile alignment for LLM service selection. The task combines coding, reasoning, and data analysis with weights of 0.50, 0.33, and 0.17 at difficulty L3. The model profiles of services A, B, and C express performance in the same capability coordinates.}
 \label{fig:routing-perspective}
\end{figure}

The central challenge is determining which past outcomes support that estimate. Holistic evaluation reveals substantial variation across scenarios and quality dimensions~\citep{liang2023holistic}, and the Benchmark Lottery shows that relative model rankings can change with the choice of evaluation tasks~\citep{dehghani2021lottery}. Dynabench further demonstrates the value of targeted challenge examples for exposing model failures~\citep{kiela2021dynabench}. Even within a topic, requests can require different combinations of capabilities. Consider identifying incorrect statistics in a Python script and repairing its implementation. Superficially similar programming examples may test routine syntax, algorithm design, or quantitative analysis, each providing different evidence about the requested repair. Instruction-following constraints introduce another source of variation, as illustrated by IFEval~\citep{zhou2023ifeval}; reasoning demands can also change the interpretation of measured performance, as shown on BIG-Bench Hard~\citep{suzgun2023bbh}. Figure~\ref{fig:routing-perspective} illustrates the resulting selection problem. A useful model description must preserve the conditions under which performance was observed, and a selector must consult that description according to the capabilities and difficulty of the incoming task.

Existing routing results establish the value of exploiting model complementarity~\citep{huang2025routereval}. RouteLLM learns pairwise preferences~\citep{ong2025routellm}, RouterDC learns query-model compatibility~\citep{chen2024routerdc}, and nearest-neighbor routing estimates quality from similar tasks~\citep{li2025knnrouting}. These advances motivate a practical question: which parts of a model's performance history are informative for a request that combines several capabilities? Global measurements provide coverage across topics, whereas related tasks preserve local context. Combining them requires distinguishing semantic resemblance from the capability demands associated with an observed outcome. This issue persists when the catalog changes. Dependencies among models, data, and system configuration complicate maintenance~\citep{sculley2015debt}; empirical studies also identify data versioning and model reuse as distinct software-engineering challenges~\citep{amershi2019engineering}. A selector should therefore accommodate updated candidate evidence while preserving a consistent rule for comparing services.

We introduce \textbf{RouteAlign} to retain the task conditions that make past performance informative for a new request. In the programming example, results on statistical reasoning and routine code editing inform different aspects of the requested repair, even when the task descriptions are semantically close. The method preserves these distinctions through profiles of performance under different capability demands and difficulty levels, together with estimates from related tasks. Within that neighborhood, each outcome's influence depends on semantic proximity and the capability demands of both the historical and current tasks. The resulting estimate reflects how a candidate's measured competence matches the request's capability mixture. A shared scorer learns to combine broad performance coverage with this local context. A common measurement schema allows the learned scoring rule to be reused when candidate evidence is refreshed after catalog changes.

Across ten datasets and 13 candidate LLMs, RouteAlign achieves the highest Sample-Avg and Dataset-Avg among the evaluated routing baselines, exceeding RouteLLM by 1.38 and 2.22 percentage points at 1.05\% higher generation cost. It also exceeds the strongest tested open Jev-style configuration by 5.22 and 6.44 points on the respective metrics at 9.7\% lower generation cost. The quality gains persist across decision backends and scoring mixtures, supporting the combination of requirement-conditioned measurements and local performance evidence. Ablations identify local evidence as the most consequential feature group. Timing studies quantify cached selection and online requirement prediction. In controlled catalog updates, refreshed measurements let the retained scorer respond to changes in available services and prices.

The paper makes three contributions:
\begin{itemize}
 \item A requirement-aware formulation of LLM service selection that separates task competence, service eligibility, and generation-cost preference. Shared capability and difficulty coordinates make the relationship between task demands and measured model performance explicit.
 \item A requirement-profile alignment mechanism that combines difficulty-conditioned model profiles with local estimates conditioned on the capabilities required by both historical and incoming tasks. A shared scorer integrates these complementary measurements across candidate services.
 \item An empirical assessment spanning established routers, open Jev-style decision backends, and direct candidate invocation. Quality, generation cost, ablations, execution overhead, and catalog updates jointly examine the effectiveness and reuse of the selection policy.
\end{itemize}

\section{Related Work}
\label{sec:related}

\subsection{Service Selection and Task-Dependent Quality}
Service selection traditionally compares functionally suitable candidates through QoS attributes and application preferences. Zeng et al.~\citep{zeng2004qos} formulate utility-based selection under user constraints and examine task-level and global allocation in composite services. Their retrospective~\citep{zeng2025retrospective} identifies stochastic outputs and unstructured data as challenges for composition involving LLMs. For an LLM invocation, task-dependent competence is itself a quantity to be estimated before candidates can be compared. RouteAlign addresses this estimation problem within an available service catalog: requirements condition the performance evidence, metadata determines admissibility, and generation cost enters the selection preference. This formulation connects service descriptions to invocation-time selection.

\subsection{LLM Routing and Resource Allocation}
Pre-generation routing estimates which model should answer a request from information available before candidate execution. Hybrid LLM~\citep{ding2024hybrid} predicts quality differences, RouteLLM~\citep{ong2025routellm} learns from pairwise preferences, and RouterDC~\citep{chen2024routerdc} uses contrastive query-model objectives. Reward-based supervision provides another source of such evidence: ZOOTER~\citep{lu2024routing} distills rewards into a routing policy, while Routoo~\citep{mohammadshahi2024routoo} couples predicted performance with cost-aware selection. RouteAlign also makes its decision before response generation, with a focus on how task requirements structure and weight the measurements supplied to that decision.

Other approaches extend the allocation decision across executions or adapt it from feedback. FrugalGPT~\citep{chen2024frugalgpt} learns model cascades, and AutoMix~\citep{aggarwal2024automix} uses answer verification to decide when to invoke a larger model. MixLLM~\citep{wang2025mixllm} formulates allocation as a contextual bandit, while BEST-Route~\citep{ding2025bestroute} jointly selects a model and the number of sampled responses. These settings introduce sequential or resource-allocation decisions beyond choosing a single service for the current invocation.

Evaluation resources separate policy design from the collection of model outcomes. RouterBench~\citep{hu2024routerbench} supplies reusable inference records, RouterEval~\citep{huang2025routereval} examines model complementarity, and LLMRouterBench~\citep{li2026llmrouterbench} supports comparison under a common routing protocol. LLMRouter~\citep{feng2026llmrouter} provides tools for implementing and deploying routing policies. Following this emphasis on comparable evidence, our evaluation uses a common held-out workload and reports task quality alongside recorded generation charges.

\subsection{Task Requirements and Model Performance Evidence}
Model documentation gives performance measurements an explicit context. Model Cards~\citep{mitchell2019modelcards} record intended uses, evaluation conditions, and disaggregated results; Datasheets for Datasets~\citep{gebru2021datasheets} document data composition, collection, and intended use. Holistic evaluation~\citep{liang2023holistic} compares language models across scenarios and metrics, while CheckList~\citep{ribeiro2020checklist} tests specific capabilities that aggregate accuracy can overlook. Benchmark design further specifies the demands under evaluation: FLORES-200 examines multilingual translation~\citep{nllb2022translation}, and WritingBench evaluates diverse writing requirements~\citep{wu2025writingbench}. WebArena measures interactive task completion~\citep{zhou2024webarena}, while HarmBench evaluates harmful behavior and robust refusal~\citep{mazeika2024harmbench}. These studies motivate preserving the task conditions underlying model measurements. RouteAlign gives those conditions a computational role by using capability demand and difficulty to organize and consult empirical evidence.

Routing methods operationalize model descriptions in several ways. GraphRouter~\citep{feng2025graphrouter} models relationships among tasks, queries, and candidates. UniRoute~\citep{jitkrittum2026universal} derives model features from predictions on representative prompts, whereas ICL-Router~\citep{wang2026iclrouter} conditions routing on in-context representations of query and model performance. AvengersPro~\citep{zhang2025avengerspro} summarizes quality and efficiency within query clusters, and nearest-neighbor routing~\citep{li2025knnrouting} estimates quality from similar tasks. These approaches motivate examining both the structure of candidate representations and the locality of supporting evidence.

Task structure further specifies what that evidence measures. IRT-Router~\citep{song2025irtrouter} relates query difficulty to model ability, and InferenceDynamics~\citep{shi2025inferencedynamics} constructs capability and knowledge profiles. SkillOrchestra~\citep{wang2026skillorchestra} derives skill-specific competence from execution experience for agent orchestration. DecoR~\citep{lv2026decor} decomposes query requirements to guide matching against historical records. RouteAlign brings these forms of evidence into a common requirement representation. Model profiles retain capability and difficulty conditions, and local estimates adjust the influence of semantically related outcomes according to task demand. The shared scorer compares candidates through features with consistent meanings across the service catalog.

\section{Requirement-Aware LLM Service Selection}
\label{sec:problem}

\subsection{Service Catalog and Selection Scope}
Consider an application with access to a catalog $\mathcal{M}_t$ of LLM services at deployment version $t$. Each candidate $m$ represents configured access to a particular model version. Its model determines task competence, while its service configuration specifies execution attributes such as context capacity, supported modalities, tool access, and availability. A request $x$, together with its supplied context, requires one eligible candidate to be selected before response generation. The decision selects one invocation target from the available catalog.

The empirical study represents each service candidate by its model configuration and recorded responses. We use routing to refer to existing methods and to implementations of this selection decision.

\subsection{Task Requirements and Model Profiles}
Task requirements specify both the competence needed for successful completion and the conditions under which a service may be used:
\begin{equation}
 R(x)=\bigl(r(x),d(x),u(x)\bigr),
 \label{eq:requirements}
\end{equation}
where $r(x)\in\mathbb{R}^{K}$ contains nonnegative capability weights with $\sum_k r_k(x)=1$, $d(x)\in\{1,\ldots,L\}$ denotes difficulty, and $u(x)$ contains execution requirements. The capability mixture describes the relative demands of the task, and difficulty specifies the level at which those capabilities are needed. Execution requirements are obtained from request metadata and deployment policy.

A profile $P_{m,t}$ associates candidate $m$ with capability measurements across difficulty levels and with its service attributes. Comparing these attributes with $u(x)$ defines the eligible set $\mathcal{F}_t(x)\subseteq\mathcal{M}_t$. For example, a candidate that lacks explicitly required tool access is excluded regardless of its estimated competence. Capability weights then govern quality estimation within the eligible set. Eligibility checks enforce known metadata requirements; missing attributes do not independently exclude a candidate.

\subsection{Selection Objective}
At a fixed deployment version, let $q(x,m)\in[0,1]$ denote expected task quality and $c(x,m)\geq0$ expected generation cost. For a nonempty eligible set, the selection objective is
\begin{equation}
 \pi_t^*(x)\in\operatorname*{arg\,max}_{m\in\mathcal{F}_t(x)}
 \bigl[q(x,m)-\lambda c(x,m)\bigr],
 \label{eq:selection-objective}
\end{equation}
where $\lambda\geq0$ specifies the application preference for cost relative to quality. This penalty expresses a soft preference, while eligibility determines which candidates may enter the comparison. The surrounding application policy handles requests for which no service is eligible.

Quality and cost must be estimated before the selected candidate executes the request. Historical evidence provides tasks $x_i$, recorded responses $y_{im}$, task scores $s_{im}\in[0,1]$, generation costs $c_{im}\geq0$, and available metadata. The challenge is to use these observations according to their relevance to the incoming requirements, rather than treating each measurement as equally informative for every task.

A model profile must distinguish performance under different task demands and support comparison across candidates. Shared requirement coordinates condition measurements on capability demand and difficulty; a shared scorer maps the resulting evidence to a common quality scale. The catalog determines which candidates are available for comparison, while the requirement representation remains independent of their identities. This distinction provides the basis for reusing task descriptions as the candidate set changes.

\section{RouteAlign: Requirement-Profile Alignment}
\label{sec:method}

\subsection{Task Requirement Profiling}
\label{sec:requirements}
\begin{figure*}[t]
 \centering
 \includegraphics[width=\textwidth]{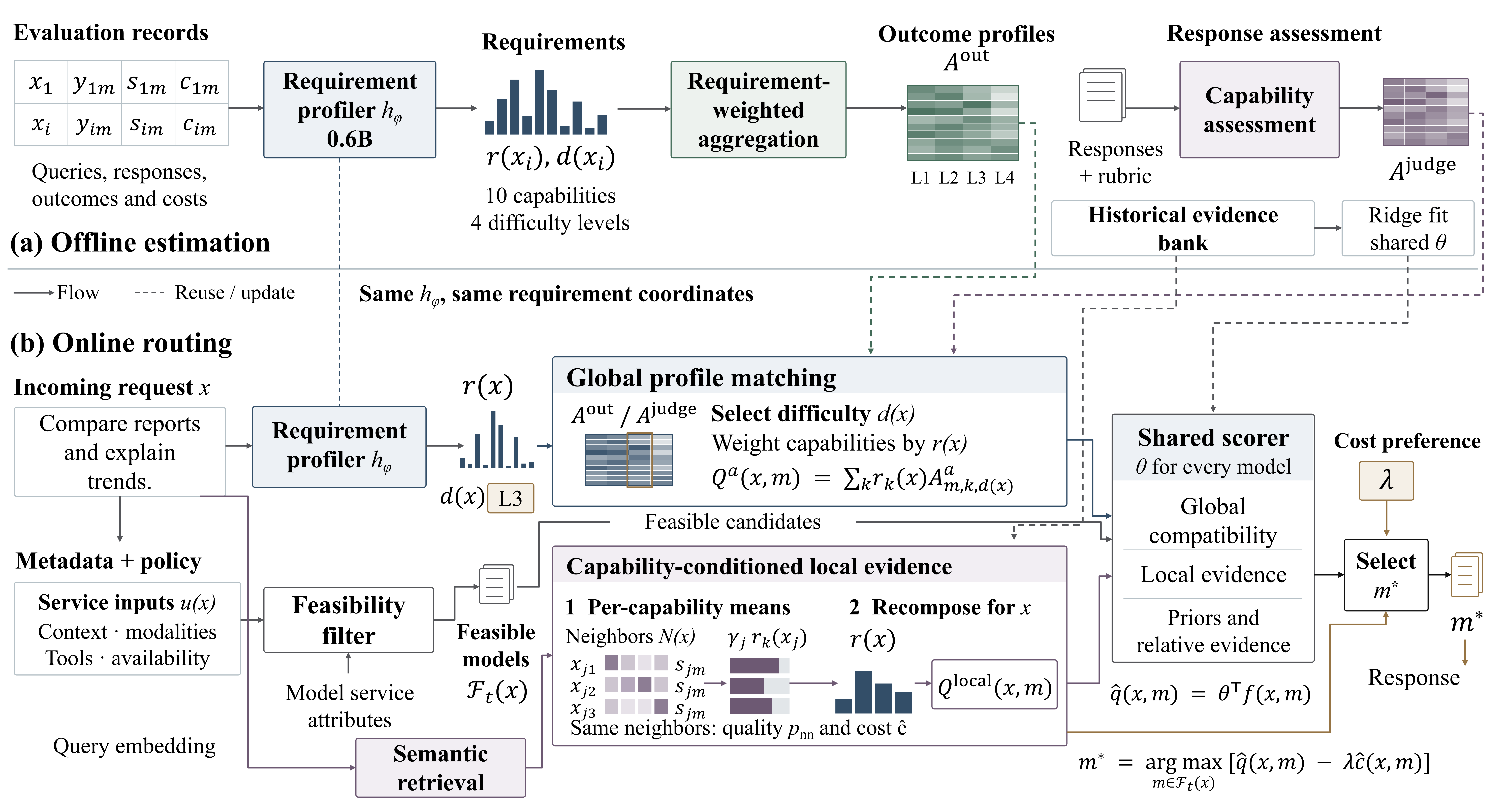}
 \caption{RouteAlign for LLM service selection. Offline preparation uses training records to construct model profiles, a historical evidence bank, and a shared scorer. The same profiler supplies requirements for outcome profiles and incoming tasks. Response assessments use separate annotations in the common schema. At invocation time, requirements align with global profiles and local evidence, metadata determines eligibility, and the cost preference shapes selection.}
 \label{fig:framework}
\end{figure*}

RouteAlign uses a shared requirement representation throughout evidence preparation and invocation-time selection, as shown in Figure~\ref{fig:framework}. Offline preparation annotates historical tasks, constructs model profiles, indexes reference tasks, and fits a shared scorer. For an incoming request, the selector predicts requirements, checks eligibility, retrieves related records, and compares the eligible candidates. Each service enters this comparison through measurements of its underlying model.

A compact profiler $h_\phi$ predicts capability weights and difficulty from the request text. Its ten capability coordinates cover task understanding, domain knowledge, reasoning, software engineering, complex information synthesis, data analysis, multilingual understanding, content generation, interactive execution, and risk control. Four difficulty levels range from direct facts and simple transformations to expert analysis and extended planning. A task can assign positive weight to several capabilities, preserving the mixture needed for completion.

We formulate profiling as instruction-conditioned generation of structured annotations, building on instruction tuning as a means of learning from task descriptions~\citep{wei2022finetuned}. Qwen3-0.6B~\citep{yang2025qwen3} receives full-parameter supervised fine-tuning (SFT) in LlamaFactory~\citep{zheng2024llamafactory}, using approximately 21,000 GPT-5.4-annotated task profiles. The source pool covers the ten capabilities, including program synthesis in HumanEval~\citep{chen2021humaneval}, numerical reasoning over financial data in FinQA~\citep{chen2021finqa}, and document reasoning in LongBench v2~\citep{bai2025longbenchv2}. Train-test overlap can distort model evaluation~\citep{lee2022deduplicating}. We therefore first exclude every source dataset present in the routing test workload, then remove residual overlap using token 3-grams: a candidate is discarded when its overlap covers at least 10\% of a compared test sample's 3-grams. Training minimizes loss on the target profile only. The same deterministic profiler annotates historical and incoming tasks; execution requirements are supplied separately through metadata. The Supplementary Material gives the source inventory, filtering rule, and training settings.

The profiler runs once per incoming request, and all candidates reuse its output. Because this output describes task demand, it can be applied to any candidate with suitable performance evidence. Candidate additions do not change the capability coordinates or difficulty levels. The same task description therefore supports both candidate measurement and selection across the catalog.

\subsection{Requirement-Conditioned Model Profiles}
\label{sec:profiles}
A global profile summarizes performance under specified capability demands and difficulty levels. For model $m$, capability $k$, and difficulty $\ell$, let $\mathcal{I}_{m,k,\ell}$ index historical tasks with an observed score, profiler-assigned difficulty $d(x_i)=\ell$, and positive capability weight $r_k(x_i)$. The outcome profile is
\begin{equation}
 A^{\mathrm{out}}_{m,k,\ell}
 =\frac{\sum_{i\in\mathcal{I}_{m,k,\ell}}r_k(x_i)s_{im}}
 {\sum_{i\in\mathcal{I}_{m,k,\ell}}r_k(x_i)}.
 \label{eq:outcome-profile}
\end{equation}
Observed successes and failures both enter this estimate, weighted by the demands of their tasks. An observation may contribute to several cells because a task can require several capabilities. Each cell describes performance on tasks that demand the corresponding capability at a specified difficulty. If a cell is unobserved, lookup uses the nearest observed difficulty level, preferring the higher level in a tie. A capability with no observed level contributes zero, and the remaining weights retain their original values.

LLM-based response evaluation provides another source of quality evidence~\citep{zheng2023judging}. RouteAlign constructs a second profile, $A^{\mathrm{judge}}$, from capability-specific assessments of recorded responses against references and task rubrics. These assessments use the common capability and difficulty schema with independently annotated requirement weights. Outcome scores remain in $[0,1]$, and response assessments remain in $[0,10]$; separate coefficients in the shared scorer account for the two channels. The assessment protocol and aggregation rule are specified in the Supplementary Material.

For an incoming task, each profile yields a global compatibility estimate:
\begin{equation}
 \begin{gathered}
 Q^a(x,m)=\sum_{k=1}^{K}r_k(x)A^a_{m,k,d(x)},\\
 a\in\{\mathrm{out},\mathrm{judge}\}.
 \end{gathered}
 \label{eq:profile-match}
\end{equation}
The predicted difficulty selects the measurement level, and the incoming capability weights determine how the selected cells contribute. Historical demands thus govern the construction of outcome measurements, while current demands govern their use. Difficulty conditioning preserves the distinction between routine and demanding applications of a capability. Weighted aggregation also allows mixed tasks to draw on several cells, instead of forcing each observation or request into one exclusive skill category.

\subsection{Capability-Conditioned Local Quality Estimation}
\label{sec:local}
Global profiles combine observations across topics, which can obscure distinctions relevant to a particular request. RouteAlign retains this local context by retrieving the $k_{\mathrm{nn}}$ nearest historical tasks in a frozen embedding space. Let $\mathcal{N}(x)$ denote the neighborhood and $\gamma_j(x)$ its normalized weights, proportional to $\exp(-\beta\delta(x,x_j))$ for cosine distance $\delta$. The scalar quality and generation-cost estimates are
\begin{equation}
 \begin{aligned}
 p_{\mathrm{nn}}(x,m)&=\sum_{j\in\mathcal{N}(x)}\gamma_j(x)s_{jm},\\
 \widehat c(x,m)&=\sum_{j\in\mathcal{N}(x)}\gamma_j(x)c_{jm}.
 \end{aligned}
 \label{eq:local-statistics}
\end{equation}
These weights reflect semantic proximity but do not distinguish the capability demands underlying neighboring outcomes. RouteAlign therefore also estimates local performance for each capability and combines those estimates according to the incoming requirements:
\begin{equation}
 \begin{aligned}
 p_k^{\mathrm{cap}}(x,m)&=
 \frac{\sum_{j\in\mathcal{N}(x)}\gamma_j(x)r_k(x_j)s_{jm}}
 {\sum_{j\in\mathcal{N}(x)}\gamma_j(x)r_k(x_j)},\\
 Q^{\mathrm{local}}(x,m)&=\sum_{k=1}^{K}r_k(x)p_k^{\mathrm{cap}}(x,m).
 \end{aligned}
 \label{eq:local-match}
\end{equation}
If the denominator is at most $10^{-6}$, the capability estimate falls back to $p_{\mathrm{nn}}(x,m)$. Neighbors need not share the incoming difficulty level. Difficulty instead conditions the global profiles and the category-by-difficulty mean used by the scorer.

The historical weight $r_k(x_j)$ controls how much a neighboring outcome informs capability $k$; the incoming weight $r_k(x)$ controls how much that capability estimate matters to the current decision. In the introductory example, a task dominated by code implementation can therefore contribute differently from one dominated by quantitative reasoning, even when both are semantically close to the request. Equations~\eqref{eq:outcome-profile} and~\eqref{eq:local-match} apply the same requirement coordinates at different scopes: the former aggregates model evidence globally, and the latter adapts evidence within the retrieved neighborhood.

The resulting influence on each neighbor can be made explicit. Let $Z_k(x)=\sum_{j\in\mathcal{N}(x)}\gamma_j(x)r_k(x_j)$. When $Z_k(x)>10^{-6}$ for every capability with positive incoming weight, Equation~\eqref{eq:local-match} can be written as
\begin{equation}
 \begin{aligned}
 Q^{\mathrm{local}}(x,m)&=\sum_{j\in\mathcal{N}(x)}\omega_j(x)s_{jm},\\
 \omega_j(x)&=\gamma_j(x)\sum_{k:r_k(x)>0}
 \frac{r_k(x)r_k(x_j)}{Z_k(x)}.
 \end{aligned}
 \label{eq:effective-weights}
\end{equation}
The effective weights are nonnegative and sum to one. Normalization by $Z_k(x)$ gives each capability a weighted mean within the neighborhood; its contribution to the final estimate is then set by the incoming requirement weight. A capability does not receive a larger share merely because it is more prevalent among the neighbors. Semantic retrieval determines which observations are available, and requirement alignment determines their relative influence. If all neighbors have the same supported capability mixture, the effective weights reduce to $\gamma_j(x)$. When a required capability has insufficient support, its portion of the estimate uses the scalar fallback.

\subsection{Shared Quality Scoring and Cost-Aware Selection}
\label{sec:fusion}
Global and local estimates summarize different aspects of performance but may share predictive information. RouteAlign fits one linear quality estimator, $\widehat q(x,m)=\theta^\top f(x,m)$, jointly across candidates. Its ten features include the two global compatibility scores, scalar and capability-conditioned local quality, the sample-category mean $b$, the category-by-difficulty mean $g$, and four local comparison features. The latter comprise local quality multiplied by candidate separation, the margin over the strongest eligible alternative, an indicator of the highest local estimate, and disagreement with the category-by-difficulty mean. Joint fitting determines the contribution of each estimate in the presence of the others.

The terms sample-category prior and category-by-difficulty prior refer to empirical means of training outcomes, selected using an auxiliary sample-category prediction. An empty category-by-difficulty stratum falls back to the category mean. Candidate separation is the maximum local quality minus the mean across eligible candidates; in the benchmark, it is normalized by the mean separation in the corresponding feature-extraction batch. Selection ties follow a deterministic candidate order, and a single eligible candidate uses zero as the reference for its margin. These definitions are shared across models, so candidate identity indexes evidence without defining a learned output class. The Supplementary Material gives the complete feature vector and specifies how missing records are handled.

Ridge regression fits the scorer to recorded training scores:
\begin{equation}
 \begin{aligned}
 \theta^*=\operatorname*{arg\,min}_{\theta}\Biggl\{
 \sum_{(i,m)\in\mathcal{P}_{\mathrm{train}}}
 \bigl(\theta^\top f(x_i,m)-s_{im}\bigr)^2
 +\alpha\|\theta\|_2^2\Biggr\}.
 \end{aligned}
 \label{eq:ridge}
\end{equation}
Training uses native feature scales and no intercept, producing an unconstrained regression score. Retrieval for a training query excludes that query itself, while global profiles and category priors aggregate the training partition. At invocation time, selection follows
\begin{equation}
 \pi_t(x)\in\operatorname*{arg\,max}_{m\in\mathcal{F}_t(x)}
 \bigl[\theta^{\top}f(x,m)-\lambda\widehat c(x,m)\bigr].
 \label{eq:routing}
\end{equation}
The fitted coefficients estimate task quality from the available evidence, and $\lambda$ determines how estimated generation expenditure affects the choice. Changing this preference changes the operating point without refitting the quality estimator. Requirement prediction and retrieval are shared across candidates; subsequent aggregation and scoring are linear in catalog size for a fixed schema and neighborhood size. Section~\ref{sec:runtime} measures the cached selection stage and a separate endpoint that includes requirement prediction.

\subsection{Updating Evidence as the Service Catalog Changes}
\label{sec:maintenance}
The shared representation separates updates to candidate measurements from fitting the comparison rule. With task annotations fixed, each outcome-profile cell depends on a total requirement weight and a weighted sum of outcomes. Reference measurements for an added candidate populate its cells without changing those of existing candidates. A model replacement requires measurements for the new version, removal changes the active set, and repricing updates service metadata and the cost records used by the local estimator. Candidate-relative features are recomputed for the eligible set.

Requirement annotations and reference embeddings remain reusable as long as the profiler, schema, and reference tasks are unchanged. Newly measured candidates can then be expressed through the existing feature definitions and evaluated with retained scoring coefficients. Section~\ref{sec:pool-experiments} tests this property by fitting on an initial catalog, freezing the coefficients, and refreshing candidate evidence after each catalog update.

\section{Empirical Evaluation}
\label{sec:experiments}

The experiments assess selection quality and generation cost, the contributions of different feature groups, selection time, and scorer reuse after catalog changes. Four research questions organize the evaluation:
\begin{itemize}
 \item \textbf{RQ1: Quality and cost.} How does RouteAlign compare with existing routers, open Jev-style decision backends, and direct candidate invocation in quality and generation expenditure?
 \item \textbf{RQ2: Feature contributions.} How do global, local, and requirement-conditioned estimates contribute to quality within the shared scorer?
 \item \textbf{RQ3: Execution overhead.} How long does selection take with cached requirements and embeddings, and what latency is observed at a separate endpoint that also predicts requirements?
 \item \textbf{RQ4: Scorer reuse.} Can coefficients fitted on an initial catalog remain effective after candidate evidence is refreshed to reflect catalog changes?
\end{itemize}

\begin{table*}[t]
 \centering
 \caption{Task quality and generation cost across ten benchmarks. S-Avg. and D-Avg. denote Sample-Avg and Dataset-Avg. Bold marks the highest quality or lowest generation cost in each column.}
 \label{tab:router-results}
 \footnotesize
 \setlength{\tabcolsep}{1.8pt}
 \resizebox{\textwidth}{!}{%
 \begin{tabular}{@{}l*{13}{r}@{}}
 \toprule
 & \multicolumn{10}{c}{Per-dataset score (\%)} & \multicolumn{2}{c}{Overall (\%)} & \\
 \cmidrule(lr){2-11}\cmidrule(lr){12-13}
 Selection policy & AIME & AH & GPQA & HLE & LCB & LMB & MMP & SQA & SWE & $\tau^2$ & S-Avg. & D-Avg. & Cost (\$) \\
 \midrule
 Random Router & 61.11 & 68.14 & 73.33 & 10.34 & 62.78 & 64.86 & 84.44 & 34.13 & 23.33 & 40.48 & 47.38 & 52.29 & 51.11 \\
 \addlinespace[1.5pt]
 KNNRouter & 61.11 & 71.46 & 58.33 & 13.73 & 63.41 & 64.86 & 82.89 & 53.93 & 28.67 & 57.14 & 55.07 & 55.55 & 34.96 \\
 SVMRouter & 77.78 & 76.99 & 55.00 & 9.72 & 58.36 & 72.97 & 83.78 & 54.01 & 16.67 & 41.67 & 53.80 & 54.69 & \textbf{6.98} \\
 MLPRouter & 72.22 & 70.80 & 61.67 & 18.52 & 67.82 & 59.46 & 82.78 & 50.69 & 24.00 & 70.24 & 55.24 & 57.82 & 42.44 \\
 \addlinespace[1.5pt]
 MM-Embed-S & 66.67 & 69.69 & 81.67 & 15.74 & 79.18 & 70.27 & 83.33 & 53.00 & 28.00 & 53.57 & 56.78 & 60.11 & 56.28 \\
 MM-Embed-L & 72.22 & 74.34 & 80.00 & 16.51 & 76.34 & 75.68 & 83.56 & 53.08 & 25.33 & 48.81 & 56.88 & 60.59 & 58.48 \\
 mmBERT-32K & 66.67 & 73.23 & 78.33 & 16.67 & 76.66 & 75.68 & 83.22 & 53.39 & 28.00 & 54.76 & 57.08 & 60.66 & 59.28 \\
 \addlinespace[1.5pt]
 GraphRouter (cost) & 55.56 & 70.13 & 70.00 & 8.33 & 64.04 & 67.57 & 83.56 & 46.84 & 17.33 & 41.67 & 51.19 & 52.50 & 7.12 \\
 GraphRouter (balance) & 77.78 & 76.99 & 70.00 & 9.88 & 58.36 & 72.97 & 84.11 & 54.01 & 17.33 & 41.67 & 54.17 & 56.31 & 10.14 \\
 GraphRouter (quality) & 66.67 & 75.22 & 58.33 & 14.97 & 64.67 & 40.54 & 83.33 & 54.01 & 30.67 & 39.29 & 55.22 & 52.77 & 146.68 \\
 \addlinespace[1.5pt]
 RouteLLM & \textbf{83.33} & 73.89 & \textbf{88.33} & 25.93 & \textbf{84.54} & 78.38 & \textbf{87.22} & 56.86 & 16.67 & 69.05 & 61.69 & 66.42 & 109.27 \\
 \addlinespace[1.5pt]
 RouterDC & \textbf{83.33} & 71.24 & \textbf{88.33} & 25.93 & \textbf{84.54} & 78.38 & \textbf{87.22} & 48.38 & 16.00 & 69.05 & 58.56 & 65.24 & 122.62 \\
 \addlinespace[1.5pt]
 AvengersPro & \textbf{83.33} & 71.24 & 76.67 & 22.84 & 77.92 & 78.38 & 85.67 & 53.54 & 26.00 & 60.71 & 58.91 & 63.63 & 61.35 \\
 \midrule
 \textbf{RouteAlign} & \textbf{83.33} & \textbf{78.10} & 81.67 & \textbf{26.08} & 83.60 & \textbf{81.08} & 86.89 & \textbf{58.47} & \textbf{33.33} & \textbf{73.81} & \textbf{63.07} & \textbf{68.64} & 110.42 \\
 \bottomrule
 \end{tabular}%
 }
 \par\smallskip
 \begin{minipage}{\linewidth}
 \footnotesize AH: ArenaHard; LCB: LiveCodeBench; LMB: LiveMathBench; MMP: MMLU-Pro; SQA: SimpleQA; SWE: SWE-bench.
 \end{minipage}
\end{table*}

\subsection{Experimental Setup}
\label{sec:experimental-setup}

\subsubsection{Datasets and Train-Test Split}
We use the requests, model responses, task scores, and generation costs recorded in LLMRouterBench~\citep{li2026llmrouterbench}. The workload comprises AIME, ArenaHard, GPQA, HLE, LiveCodeBench, LiveMathBench, MMLU-Pro, SimpleQA, SWE-bench, and $\tau^2$-bench. Partitioning by dataset and prompt with seed 42 and a 70\% training fraction produces 8,710 training requests and 3,738 test requests. All responses to a request remain in the same partition. Model measurement, retrieval references, category priors, and scorer fitting use training records. A policy chooses a candidate before accessing its recorded test outcome.

The 13 model configurations represent alternative LLM services whose recorded outcomes permit comparison on a shared workload. Direct-candidate results use available records: GPT-5 covers the full test workload, whereas GPT-5-Chat lacks $\tau^2$-bench records. Dense retrieval features and regression targets use zero for absent records; global outcome profiles aggregate observed records only. RouteAlign selects an observed response for every test request. Dataset sizes, candidate coverage, and selection counts are reported in the Supplementary Material.

\subsubsection{Baseline Policies and Implementations}
The comparison includes preference-based RouteLLM~\citep{ong2025routellm}, contrastive RouterDC~\citep{chen2024routerdc}, three utility configurations of GraphRouter~\citep{feng2025graphrouter}, and cluster-based AvengersPro~\citep{zhang2025avengerspro}. KNNRouter, SVMRouter, and MLPRouter provide classification-based comparisons~\citep{feng2026llmrouter}. We also evaluate three embedding-based retrieval policies, uniform random selection, and direct invocation of individual candidates. This set covers several ways of learning or aggregating historical performance evidence.

RouteLLM uses a multi-candidate matrix-factorization adaptation, and AvengersPro uses the benchmark implementation of cluster-based rank aggregation. Results refer to the implementations and training budgets detailed in the Supplementary Material. In particular, classification-based KNNRouter predicts a winning candidate label, whereas the retrieval-based policies estimate performance separately for each candidate.

\subsubsection{Evaluation Metrics and Configuration}
Sample-Avg weights requests equally, while Dataset-Avg weights dataset means equally. Both preserve graded task scores and are reported as percentages. Generation cost sums recorded candidate charges in US dollars, excluding requirement prediction, selector execution, offline profiling, and scorer fitting. Selector runtime is measured separately. Quality-cost comparisons use each policy's observed expenditure at its specified operating point.

RouteAlign uses the profiler in Section~\ref{sec:requirements} and sentence embeddings for semantic retrieval~\citep{reimers2019sentence}. The retrieval encoder is the frozen \texttt{all-MiniLM-L6-v2} checkpoint with 384-dimensional outputs. Defaults are $k_{\mathrm{nn}}=150$, $\beta=5$, ridge penalty $\alpha=1$, and cost preference $\lambda=0.4$. Feature ablations retain the partition and definitions of the remaining features and refit the scorer.

\subsection{Selection Quality and Generation Cost}
\label{sec:fixed-pool-results}
\label{sec:quality-cost}

\subsubsection{Aggregate and Per-Dataset Quality}
RouteAlign achieves the highest aggregate quality among the evaluated policies in the fixed catalog, with 63.07\% Sample-Avg and 68.64\% Dataset-Avg (Table~\ref{tab:router-results}). The strongest competing router, RouteLLM, obtains 61.69\% and 66.42\%, respectively. The respective gains of 1.38 and 2.22 percentage points show improvement at the workload level and under equal weighting of datasets.

Per-dataset scores reveal how the advantage varies across task groups. Among the policies in Table~\ref{tab:router-results}, RouteAlign ranks first, including ties, on seven datasets, while RouteLLM and RouterDC achieve higher scores on GPQA, LiveCodeBench, and MMLU-Pro. The largest gain over RouteLLM occurs on SWE-bench, from 16.67\% to 33.33\%. This distribution of gains motivates examining the evidence used to select candidates for different task demands.

\begin{table}[t]
 \centering
 \captionsetup{width=0.70\linewidth}
 \caption{Comparison with direct invocation of individual candidates. Scores (\%) and generation costs (US dollars) use available records. GPT-5 covers all test requests; GPT-5-Chat excludes $\tau^2$-bench.}
 \label{tab:single-model-results}
 \footnotesize
 \setlength{\tabcolsep}{2.3pt}
 \begin{tabular*}{0.70\linewidth}{@{\extracolsep{\fill}}lrrr@{}}
 \toprule
 & \multicolumn{2}{c}{Quality (\%)} & \\
 \cmidrule(lr){2-3}
 Candidate or policy & S-Avg. & D-Avg. & Cost (\$) \\
 \midrule
 Intern-S1 & 37.74 & 43.03 & 13.85 \\
 DeepSeek V3-0324 & 40.41 & 42.60 & 4.78 \\
 DeepSeek V3.1-Terminus & 44.18 & 51.72 & 9.43 \\
 DeepSeek R1-0528 & 47.85 & 55.57 & 60.75 \\
 Qwen3-235B & 53.81 & 54.70 & 6.98 \\
 Qwen3-235B Thinking & 53.13 & 56.27 & 36.84 \\
 Kimi K2-0905 & 45.09 & 54.76 & 14.53 \\
 GLM-4.6 & 44.65 & 55.48 & 61.39 \\
 GPT-5-Chat & 46.89 & 50.83 & 17.96 \\
 Gemini 2.5 Flash & 42.64 & 45.74 & 32.97 \\
 Claude Sonnet 4 & 38.41 & 46.47 & 70.28 \\
 Gemini 2.5 Pro & 59.18 & 59.55 & 233.97 \\
 GPT-5 & 58.37 & 65.11 & 124.82 \\
 \midrule
 AVG LLM & 47.10 & 52.45 & 52.96 \\
 Random Router & 47.38 & 52.29 & 51.11 \\
 \midrule
 \textbf{RouteAlign} & \textbf{63.07} & \textbf{68.64} & 110.42 \\
 \bottomrule
 \end{tabular*}
\end{table}

Table~\ref{tab:single-model-results} tests the value of selection against committing to one candidate. GPT-5 has the highest individual Dataset-Avg and covers the same 3,738 requests as RouteAlign. Against this default, RouteAlign gains 4.70 Sample-Avg points and 3.53 Dataset-Avg points. It also exceeds the highest individual Sample-Avg, achieved by Gemini 2.5 Pro, by 3.89 points. RouteAlign assigns 1,852 requests to GPT-5 and the remaining 1,886 to other candidates. Its advantage over a strong default is therefore obtained through selective use of the catalog. Section~\ref{sec:ablations} examines the feature contributions underlying these decisions.

\subsubsection{Quality and Generation Cost Comparisons}
\begin{figure*}[t]
 \centering
 \includegraphics[width=0.95\textwidth]{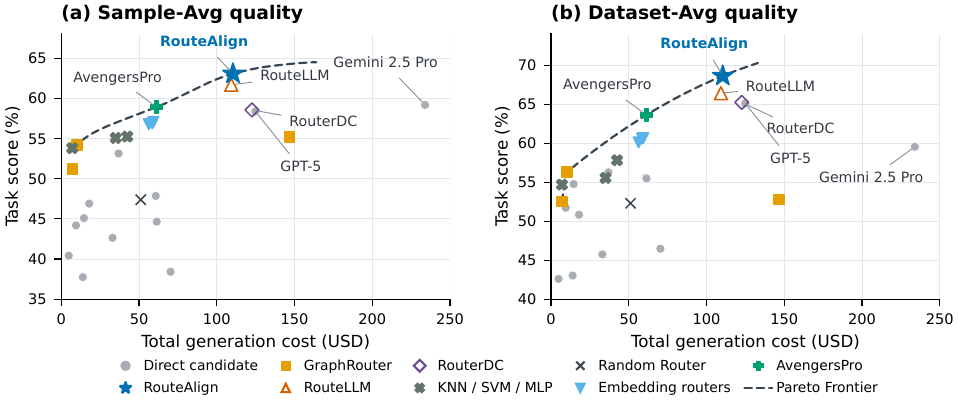}
 \caption{Task quality versus generation cost for the fixed-catalog configurations. Grey circles denote direct candidate invocation; colored markers denote selection policies. Dashed curves are schematic guides to the Pareto frontiers. Cost includes answer generation only; direct-candidate summaries use available records.}
 \label{fig:quality-cost}
\end{figure*}

Figure~\ref{fig:quality-cost} shows the empirical quality-cost frontiers under both aggregate metrics. RouteAlign is Pareto nondominated in each panel and achieves the highest task quality among the plotted configurations. Its generation cost of \$110.42 exceeds RouteLLM's \$109.27 by \$1.15, or 1.05\%, for gains of 1.38 Sample-Avg and 2.22 Dataset-Avg points. RouteAlign thus extends the frontier toward higher quality with a modest increase in expenditure.

RouteAlign also dominates RouterDC and direct GPT-5 invocation in both panels, achieving higher quality at lower generation cost. Against GPT-5, it reduces expenditure from \$124.82 to \$110.42, a saving of 11.5\%.

\subsection{Comparison with Jev-Style Decision Models}
\label{sec:jev}

This comparison examines whether a general decision model supplied with task and model descriptions can replace the trained selection policy. We evaluate nine locally deployed, open Jev-style backends spanning 0.8B to 9B: three Kev sizes, two Open-Jev sizes, two Decider sizes, SemIf, and OpenJev-mjdileep. Each exposes typed choices and scores through a common local adapter.

Every configuration uses the same 3,738 requests, 13-model catalog, and recorded outcomes. Each backend predicts requirements and rates candidate profile summaries, retaining the adapter's metadata filtering and capability-tier matching. The backends receive no additional routing-task fine-tuning. We vary the decision-score weight $\rho\in\{0,0.5,1\}$ in a mixture with explicit requirement-profile compatibility. At $\rho=0$, the backend still supplies the requirements and candidate set, while compatibility determines the candidate scores. The comparison evaluates complete selection policies: the Jev-style variants use backend predictions and supplied profiles, and RouteAlign combines its supervised profiler with local outcome evidence and a fitted scorer.

\begin{table}[t]
 \centering
 \captionsetup{width=0.70\linewidth}
 \caption{Open Jev-style backends with direct decision scoring ($\rho=1$), plus the strongest configuration across all three mixtures. Cost covers recorded candidate generation. Bold marks the highest aggregate quality.}
 \label{tab:jev-comparison}
 \footnotesize
 \setlength{\tabcolsep}{2.5pt}
 \begin{tabular*}{0.70\linewidth}{@{\extracolsep{\fill}}lrrr@{}}
 \toprule
 Decision backend & D-Avg. & S-Avg. & Cost (\$) \\
 \midrule
 Kev-0.8B & 51.83 & 49.65 & 39.33 \\
 OpenJev-mjdileep-0.8B & 47.44 & 45.85 & 52.57 \\
 Open-Jev-2B & 48.93 & 48.86 & 39.52 \\
 Decider-2B & 55.00 & 47.11 & 95.63 \\
 SemIf-4B & 53.43 & 51.82 & 29.32 \\
 Kev-4B & 61.43 & 55.43 & 102.97 \\
 Decider-4B & 60.61 & 57.08 & 109.32 \\
 Open-Jev-9B & 53.44 & 47.61 & 44.66 \\
 Kev-9B & 60.50 & 56.51 & 111.32 \\
 \midrule
 Decider-4B ($\rho=0$) & 62.20 & 57.85 & 122.28 \\
 \textbf{RouteAlign} & \textbf{68.64} & \textbf{63.07} & 110.42 \\
 \bottomrule
 \end{tabular*}
\end{table}

\begin{figure*}[t]
 \centering
 \includegraphics[width=\textwidth]{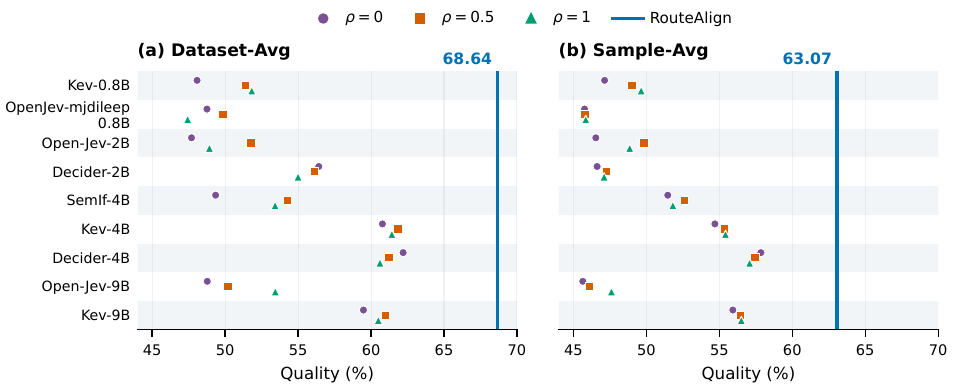}
 \caption{Aggregate quality of nine open Jev-style backends under three scoring mixtures. Each row uses one backend for requirement prediction; $\rho$ weights its decision score, with explicit requirement-profile compatibility weighted by $1-\rho$. Vertical lines mark full RouteAlign. All 27 configurations use the same held-out requests and catalog.}
 \label{fig:jev-comparison}
\end{figure*}

RouteAlign leads all 27 configurations on both aggregate metrics (Figure~\ref{fig:jev-comparison}). The strongest configuration, Decider-4B with $\rho=0$, reaches 62.20\% Dataset-Avg and 57.85\% Sample-Avg. RouteAlign gains 6.44 and 5.22 percentage points while reducing generation cost by 9.7\% (Table~\ref{tab:jev-comparison}). Under direct decision scoring, Kev-4B has the highest Dataset-Avg and Decider-4B the highest Sample-Avg. Against the latter at similar expenditure, RouteAlign gains 8.03 and 5.99 points for 1.0\% additional generation cost. Its advantage persists across all three mixtures, supporting the effectiveness of a policy trained to connect task requirements with global and local performance evidence. Complete mixture results, per-dataset scores, backend configurations, and project links appear together in the Supplementary Material.

\subsection{Ablation Analysis}
\label{sec:ablations}

\subsubsection{Ablation Protocol}
RQ2 examines the fitted combination through the complete set of feature ablations in Table~\ref{tab:full-ablation}. Each ablation removes specified feature columns and refits ridge regression on the same training partition using leave-one-out retrieval features. The encoder and definitions of retained features remain fixed. Refitting permits compensation among estimates, so a measured change reflects the contribution of the removed columns conditional on the retained evidence. The cost-preference ablation instead retains the fitted quality estimator and sets $\lambda=0$.

Removing all local quality evidence deletes scalar neighborhood quality, capability-conditioned local compatibility, and four candidate-comparison features, while retaining the generation-cost estimate. The requirement-conditioned group contains $Q^{\mathrm{out}}$, $Q^{\mathrm{judge}}$, the category-by-difficulty prior $g$, and $Q^{\mathrm{local}}$. Global profile compatibility contains the two global scores. The Supplementary Material lists the removed columns and clarifies which derived features remain when an individual input is deleted.

\begin{table*}[t]
 \centering
 \captionsetup{width=0.85\linewidth}
 \caption{Full ablation results for RouteAlign. $\Delta$ denotes the change from full RouteAlign in percentage points. Bold identifies the reference configuration.}
 \label{tab:full-ablation}
 \small
 \setlength{\tabcolsep}{4pt}
 \begin{tabular*}{0.85\linewidth}{@{\extracolsep{\fill}}lrrrrr@{}}
 \toprule
 Removed evidence or preference & D-Avg. & $\Delta$D & S-Avg. & $\Delta$S & Cost (\$) \\
 \midrule
 \textbf{Full RouteAlign} & 68.64 & 0.00 & 63.07 & 0.00 & 110.42 \\
 \midrule
 All local quality evidence & 65.52 & $-3.12$ & 61.41 & $-1.66$ & 106.24 \\
 Scalar neighborhood estimate & 68.03 & $-0.61$ & 62.88 & $-0.19$ & 110.41 \\
 Sample-category prior & 68.05 & $-0.59$ & 62.93 & $-0.14$ & 111.85 \\
 Local capability conditioning & 68.12 & $-0.52$ & 63.08 & $+0.01$ & 110.96 \\
 Relative local evidence & 68.18 & $-0.46$ & 63.22 & $+0.15$ & 109.42 \\
 Response assessment compatibility & 68.26 & $-0.38$ & 63.06 & $-0.01$ & 110.00 \\
 Cost penalty ($\lambda=0$) & 68.27 & $-0.37$ & 63.32 & $+0.25$ & 124.82 \\
 Requirement-conditioned scores & 68.39 & $-0.25$ & 63.15 & $+0.08$ & 114.73 \\
 Global profile compatibility & 68.53 & $-0.11$ & 62.99 & $-0.08$ & 109.86 \\
 Category and difficulty prior & 68.83 & $+0.19$ & 62.92 & $-0.15$ & 109.02 \\
 Outcome compatibility & 68.64 & $0.00$ & 63.07 & $0.00$ & 110.59 \\
 \bottomrule
 \end{tabular*}
\end{table*}

\subsubsection{Local Evidence versus Category Means}
The largest quality loss follows removal of all local quality evidence: 3.12 Dataset-Avg points and 1.66 Sample-Avg points. Even after refitting, global profiles and category priors do not replace the information supplied by related task outcomes. Removing the scalar neighborhood estimate alone costs 0.61 Dataset-Avg points and 0.19 Sample-Avg points. Removing the four relative local features lowers Dataset-Avg by 0.46 points but raises Sample-Avg by 0.15 points. Local evidence is therefore consequential at the group level, while its relative features affect quality differently under the two task-weighting conventions.

Selection by category means provides a complementary check on whether coarse task groupings are sufficient. At similar observed generation expenditure, selection by the category-by-difficulty mean trails RouteAlign by 2.22 Dataset-Avg points and 1.62 Sample-Avg points. The complete policy therefore improves on selecting the candidate with the highest mean score in the predicted category and difficulty group. The supplementary comparison reports the complete decision rules and their generation costs.

\subsubsection{Requirement Conditioning and Feature Overlap}
Local capability conditioning contributes most clearly under equal dataset weighting. Removing $Q^{\mathrm{local}}$ while retaining scalar neighborhood evidence lowers Dataset-Avg by 0.52 points, whereas Sample-Avg changes from 63.07\% to 63.08\%. This tests the explicit contribution of $Q^{\mathrm{local}}$ to the refitted scorer. Removing response-assessment compatibility lowers Dataset-Avg by 0.38 points and Sample-Avg by 0.01 points; removing outcome compatibility alone leaves both rounded averages unchanged.

Grouped removals reveal overlap among the estimates. Removing the four requirement-conditioned scores lowers Dataset-Avg by 0.25 points; removing the two global scores lowers it by 0.11 points. The sample-category prior improves both metrics, while the category-by-difficulty prior has mixed effects. Because refitting redistributes weight across retained features, these changes characterize the fitted combination. Taken together, the ablations identify local context as the most consequential evidence group and show that local capability conditioning contributes under equal dataset weighting.

\subsubsection{Effect of the Cost Preference}
Setting $\lambda=0$ increases generation expenditure from \$110.42 to \$124.82, lowers Dataset-Avg by 0.37 points, and raises Sample-Avg by 0.25 points. The default preference reduces generation cost by 11.5\% relative to this setting while changing the distribution of quality across tasks. Because the quality estimator is held fixed, the difference isolates the effect of applying the cost preference to the learned scores. The result shows that the preference changes candidate choices without refitting the scorer; its appropriate value depends on the workload and desired quality-cost balance.

\subsection{Selection Time and Endpoint Latency}
\label{sec:runtime}

\subsubsection{Selection with Cached Requirements and Embeddings}
With requirement profiles and embeddings cached, RouteAlign selects candidates for 3,738 requests in approximately 119 seconds, averaging 31.9 milliseconds per request (Table~\ref{tab:router-runtime}). Its selection stage is 8.9 milliseconds longer than RouteLLM's and shorter than the 45.20 milliseconds reported for AvengersPro. RouterDC and the embedding-based retrieval policies achieve lower selection times. Together with Section~\ref{sec:quality-cost}, these measurements locate RouteAlign's quality gains within the execution costs of the evaluated policies.

\begin{table}[t]
 \centering
 \captionsetup{width=0.64\linewidth}
 \caption{Selection-stage runtime over 3,738 test requests. RouteAlign uses cached requirement profiles and embeddings. Candidate answer generation is excluded.}
 \label{tab:router-runtime}
 \footnotesize
 \setlength{\tabcolsep}{3pt}
 \begin{tabular*}{0.64\linewidth}{@{\extracolsep{\fill}}lrr@{}}
 \toprule
 Selection policy & Total (s) & Per query (ms) \\
 \midrule
 Random Router & 0.0011 & 0.0003 \\
 \addlinespace[1.5pt]
 KNNRouter & 76.00 & 20.33 \\
 SVMRouter & 83.86 & 22.43 \\
 MLPRouter & 75.56 & 20.21 \\
 \addlinespace[1.5pt]
 MM-Embed-S & 4.23 & 1.13 \\
 MM-Embed-L & 9.42 & 2.52 \\
 mmBERT-32K & 9.23 & 2.47 \\
 \addlinespace[1.5pt]
 GraphRouter (cost) & 9.54 & 2.55 \\
 GraphRouter (balance) & 9.67 & 2.59 \\
 GraphRouter (quality) & 9.43 & 2.52 \\
 \addlinespace[1.5pt]
 RouteLLM & 85.97 & 23.00 \\
 \addlinespace[1.5pt]
 RouterDC & 42.33 & 11.32 \\
 \addlinespace[1.5pt]
 AvengersPro & 168.94 & 45.20 \\
 \midrule
 \textbf{RouteAlign} & 119 & 31.9 \\
 \bottomrule
 \end{tabular*}
\end{table}

The measured interval includes retrieval, feature construction, and candidate scoring. Requirement prediction, embedding construction, offline preparation, and candidate answer generation are excluded.

\subsubsection{Endpoint Latency with Online Requirement Prediction}
A separate endpoint executes the 0.6B profiler and selection with a 500-request reference bank. At concurrency 32, it processes 96 requests at 51.2 requests per second, with median latency of 0.60 seconds and 95th-percentile latency of 0.74 seconds. No errors or fallbacks occur. Timing ends with candidate selection. For request-level execution, this endpoint uses unnormalized candidate separation and alternative reference means when category metadata is absent. It characterizes latency with online requirement prediction; Table~\ref{tab:router-runtime} measures the cached selection stage of the fixed-catalog quality configuration.

\subsection{Scorer Reuse under Service Catalog Changes}
\label{sec:pool-experiments}

\subsubsection{Catalog Update Protocol}
RQ4 tests whether updating candidate measurements is sufficient to retain effective selection with a fixed scorer, as proposed in Section~\ref{sec:maintenance}. T0 contains seven candidates: Intern-S1, DeepSeek V3-0324, Qwen3-235B-2507, Kimi K2-0905, GLM-4.6, GPT-5-Chat, and Gemini 2.5 Flash. T1 through T5 add candidates, and T6 replaces a model version. T7 and T8 independently branch from T6, respectively removing GPT-5 and halving the price of Gemini 2.5 Flash. Table~\ref{tab:evolution-stages} lists the events and candidate counts. These controlled updates change candidate availability, performance evidence, or generation prices.

\begin{table}[t]
 \centering
 \captionsetup{width=0.88\linewidth}
 \caption{Service catalog update protocol. Each configuration is defined by its parent and update event; candidates are indexed by their model configurations.}
 \label{tab:evolution-stages}
 \small
 \setlength{\tabcolsep}{3pt}
 \begin{tabular*}{0.88\linewidth}{@{\extracolsep{\fill}}ll>{\raggedright\arraybackslash}p{0.54\linewidth}r@{}}
 \toprule
 Stage & Parent & Event & Candidates \\
 \midrule
 T0 & n/a & Initial candidate pool & 7 \\
 T1 & T0 & Add DeepSeek R1-0528 & 8 \\
 T2 & T1 & Add Qwen3-235B Thinking-2507 & 9 \\
 T3 & T2 & Add Claude Sonnet 4 & 10 \\
 T4 & T3 & Add Gemini 2.5 Pro & 11 \\
 T5 & T4 & Add GPT-5 & 12 \\
 T6 & T5 & Replace DeepSeek V3-0324 with V3.1-Terminus & 12 \\
 \midrule
 T7 & T6 & Remove GPT-5 & 11 \\
 T8 & T6 & Gemini 2.5 Flash price multiplied by $0.5$ & 12 \\
 \bottomrule
 \end{tabular*}
\end{table}

RouteAlign-Frozen fits its coefficients on the seven T0 candidates and retains them for all subsequent configurations. Newly admitted candidates are evaluated on training-partition reference tasks to obtain their measurements and profiles. Requirement predictions remain fixed, while candidate evidence and relative comparison features are refreshed. This protocol tests whether one learned comparison rule remains effective as measured candidates enter or leave the catalog.

RouteLLM and RouterDC retain checkpoints trained on all 13 candidates and mask unavailable outputs. Their fitting budgets are 50,000 preference pairs and 6,000 requests, respectively. AvengersPro recomputes clusters and candidate rankings from training evidence for each catalog.

\begin{table*}[t]
 \centering
 \caption{Selection quality under controlled catalog changes. Entries are Dataset-Avg (\%); Mean weights the nine configurations equally. T7 and T8 independently branch from T6. Bold marks the highest score in each column.}
 \label{tab:pool-evolution}
 \small
 \setlength{\tabcolsep}{3pt}
 \resizebox{\textwidth}{!}{%
 \begin{tabular}{@{}l*{10}{r}@{}}
 \toprule
 & \multicolumn{7}{c}{Main sequence} & \multicolumn{2}{c}{T6 branches} & \\
 \cmidrule(lr){2-8}\cmidrule(lr){9-10}
 Selection policy & T0 & T1 & T2 & T3 & T4 & T5 & T6 & T7 & T8 & Mean \\
 \midrule
 RouteLLM & 54.70 & 54.70 & 54.70 & 54.70 & 54.70 & 66.42 & 66.42 & 54.70 & 66.42 & 58.61 \\
 RouterDC & 55.79 & 55.56 & 58.69 & 58.69 & 58.69 & 65.24 & 65.24 & 58.69 & 65.24 & 60.20 \\
 AvengersPro & 52.94 & 52.69 & 54.52 & 58.52 & 57.81 & 60.49 & 61.88 & 60.58 & 64.72 & 58.24 \\
 \midrule
 \textbf{RouteAlign-Frozen} & \textbf{60.36} & \textbf{61.33} & \textbf{62.26} & \textbf{63.48} & \textbf{64.18} & \textbf{68.55} & \textbf{68.60} & \textbf{64.08} & \textbf{68.64} & \textbf{64.61} \\
 \bottomrule
 \end{tabular}%
 }
\end{table*}

\subsubsection{Service Addition and Model Replacement}
RouteAlign-Frozen achieves the highest Dataset-Avg in all nine configurations (Table~\ref{tab:pool-evolution}). Its equally weighted mean is 64.61\%, compared with 60.20\% for RouterDC, the strongest baseline by this mean. The difference is 4.41 percentage points before rounding. A supplementary figure plots the addition and replacement sequence separately from the removal and repricing branches.

Along the main sequence, RouteAlign-Frozen rises from 60.36\% at T0 to 64.18\% at T4. Adding GPT-5 at T5 contributes a further 4.37 points, reaching 68.55\%; the replacement at T6 yields 68.60\%. The coefficients remain unchanged throughout. The retained scorer can therefore use newly supplied model measurements within the existing requirement coordinates and feature definitions, providing empirical support for the separation between evidence updates and parameter fitting.

\subsubsection{Service Removal and Repricing}
The two branches distinguish a change in available competence from a change in cost evidence. Removing GPT-5 at T7 lowers RouteAlign-Frozen by 4.52 points to 64.08\%, quantifying the loss from withdrawing this candidate. The retained scorer still achieves the highest Dataset-Avg among the evaluated policies on the remaining catalog. Halving the price of Gemini 2.5 Flash at T8 yields 68.64\%, an increase of 0.04 points relative to T6, with GPT-5 retained in the catalog.

Across these updates, refreshed reference measurements allow the retained scorer to compare new model versions, select from a reduced catalog, and respond to revised prices. RouteAlign-Frozen achieves the highest quality in every configuration using coefficients fitted on the initial seven candidates.

\section{Conclusion}
\label{sec:conclusion}
RouteAlign makes task requirements the basis for connecting measured model competence to service selection. Historical capability demands determine how outcomes form global profiles and local estimates; incoming demands determine how the shared scorer uses this evidence to assess candidate suitability. The resulting policy attains the highest quality on both empirical Pareto frontiers and exceeds the strongest tested Jev-style configuration by 6.44 Dataset-Avg points at 9.7\% lower generation cost. The largest ablation loss follows removal of local evidence, establishing its central role in the fitted policy. RouteAlign-Frozen also leads across the tested catalog changes, showing that refreshed candidate measurements remain useful with the original scoring coefficients. These findings support requirement alignment as a consistent basis for combining global measurements, related task outcomes, and a reusable comparison rule.

\bibliographystyle{IEEEtranN}
\bibliography{IEEEtranBSTcontrol,main}

\beginappendix

\section{Task Requirement Schema and Profiler Training}
\label{supp:schema}

\subsection{Capability Definitions}
The requirement schema expresses task demand through capability weights and a difficulty level. Its ten nonnegative weights sum to one, allowing several capabilities to contribute to the same request. Task profiles and model measurements use the following capabilities in a fixed order.

\begin{enumerate}
 \item[\textbf{C1.}] \textbf{Task understanding and constraint adherence.} Identifying explicit and implicit task goals, interpreting instructions and conversational intent, and satisfying substantive constraints, priorities, procedures, and output requirements.
 \item[\textbf{C2.}] \textbf{Knowledge understanding and domain-specific question answering.} Understanding and applying factual, commonsense, encyclopedic, and domain-specific knowledge to answer questions accurately.
 \item[\textbf{C3.}] \textbf{Reasoning and quantitative problem solving.} Drawing deductive or inductive inferences, performing quantitative calculations, constructing proofs, planning, analyzing causal relations, and decomposing complex problems.
 \item[\textbf{C4.}] \textbf{Programming and software engineering.} Generating, understanding, debugging, and testing code, implementing algorithms, and reasoning about software structure, repositories, and programming interfaces.
 \item[\textbf{C5.}] \textbf{Complex information understanding and long-context synthesis.} Locating and reconciling evidence in complex documents, long texts, or extended conversational context, with emphasis on integration across passages and sources.
 \item[\textbf{C6.}] \textbf{Data analysis and structured information understanding.} Understanding tables, statistics, and other structured representations; extracting, comparing, calculating, interpreting, or transforming their contents.
 \item[\textbf{C7.}] \textbf{Multilingual understanding, translation, and localization.} Understanding and expressing content across languages, translating meaning, and adapting wording to linguistic and cultural contexts.
 \item[\textbf{C8.}] \textbf{Content generation and expression control.} Generating, summarizing, rewriting, or expanding content while controlling organization, style, tone, and audience suitability.
 \item[\textbf{C9.}] \textbf{Tool use and interactive task execution.} Executing tasks through retrieval, code execution, function calls, or interaction with external systems, including coordination across multiple steps.
 \item[\textbf{C10.}] \textbf{Safety, reliability, and risk control.} Recognizing safety-sensitive requirements, handling consequential advice reliably, protecting privacy, and maintaining appropriate response and refusal boundaries.
\end{enumerate}

\subsection{Difficulty Levels}
Difficulty specifies the required level of knowledge, reasoning, constraint interaction, and execution complexity. L1 comprises direct facts and simple transformations; L2 covers bounded reasoning and standard knowledge application; L3 includes substantial multistep reasoning, complex code analysis, or synthesis; and L4 covers expert analysis, extended planning, or tasks that combine several demanding capabilities. These levels apply to the task as a whole, alongside its capability mixture. The introductory illustration in the main paper abbreviates C4, C3, and C6 as coding, reasoning, and data analysis, respectively.

\subsection{Training Sources and Test-Set Separation}
\label{supp:profiler-data}
The profiler training corpus is constructed for coverage of the ten capability dimensions, with GPT-5.4 providing structured requirement annotations. Source selection begins with 33 datasets. We exclude all nine sources that also occur in the routing test workload: AIME, LiveMathBench, LiveCodeBench, GPQA, HLE, ArenaHard, MMLU-Pro, SWE-bench, and $\tau^2$-bench. Exclusion applies to each source dataset as a whole. The remaining 24 sources are organized in Table~\ref{tab:profiler-sources}. This grouping guides sampling coverage; each retained task receives its own capability mixture and difficulty annotation.

\begin{table*}[t]
 \centering
 \caption{Sources retained for profiler SFT after excluding datasets in the routing test workload. The groups indicate sampling coverage; each task receives its own capability mixture and difficulty annotation.}
 \label{tab:profiler-sources}
 \small
 \setlength{\tabcolsep}{4pt}
 \begin{tabular*}{\textwidth}{@{\extracolsep{\fill}}lp{0.77\textwidth}@{}}
 \toprule
 Capability coverage & Retained sources \\
 \midrule
 C1: Task understanding & IFEval~\citep{zhou2023ifeval}, EmoryNLP, MELD \\
 C2: Domain knowledge & MedQA, C-Eval \\
 C3: Reasoning & MATH-500, MathBench, KOR-Bench, K\&K (\texttt{kandk}), BBH~\citep{suzgun2023bbh} \\
 C4: Software engineering & MBPP, HumanEval~\citep{chen2021humaneval} \\
 C5: Long-context synthesis & LongBench v2~\citep{bai2025longbenchv2} \\
 C6: Structured information & FinQA~\citep{chen2021finqa} \\
 C7: Multilingual tasks & INCLUDE, Global-MMLU, FLORES-200~\citep{nllb2022translation} \\
 C8: Content generation & WritingBench~\citep{wu2025writingbench} \\
 C9: Interactive execution & GAIA, WebArena~\citep{zhou2024webarena} \\
 C10: Risk control & HarmBench~\citep{mazeika2024harmbench}, StrongREJECT, XSTest, BBQ~\citep{parrish2022bbq} \\
 \bottomrule
 \end{tabular*}
\end{table*}

After source-level exclusion, token-level 3-gram comparison checks the remaining candidate samples against routing test samples. This additional filtering addresses residual text overlap, which can distort language-model evaluation~\citep{lee2022deduplicating}. Let $G_3(v)$ be the collection of token 3-grams extracted from sample $v$. For a candidate training sample $v$ and a test sample $t$, the overlap ratio is
\begin{equation}
 O_3(v,t)=\frac{|G_3(v)\cap G_3(t)|}{|G_3(t)|}.
 \label{eq:profiler-overlap}
\end{equation}
A candidate is filtered if $O_3(v,t)\geq0.10$ for any compared test sample. The denominator is the 3-gram count of the test sample, so the threshold measures how much of that sample is covered by the candidate. This stage found little residual overlap after dataset exclusion. The resulting supervision comprises approximately 21,000 task-profile pairs.

\subsection{Full-Parameter Supervised Fine-Tuning}
\label{supp:profiler-sft}
We fine-tune Qwen3-0.6B~\citep{yang2025qwen3} with LlamaFactory~\citep{zheng2024llamafactory}. Each input contains the profiling instructions and the request with its supplied context. The target is a structured record containing all ten capability weights and one difficulty level. Annotation instructions assess the capabilities necessary for successful completion from the request and supplied context, with reference answers, candidate identities, and source-dataset labels excluded from the annotation input. The training objective is conditional token cross-entropy over the target profile; instruction and request tokens are masked from the loss. The supervised output thus describes the demands of the task.

Training updates all model parameters for one epoch on two NVIDIA A800 GPUs using bfloat16 and DeepSpeed ZeRO-3. The learning rate is $10^{-5}$ with cosine decay and a 0.1 warmup ratio. A per-device batch size of one and 16 gradient-accumulation steps give an effective batch size of 32. The sequence limit is 12,000 tokens, and the \texttt{qwen3\_nothink} template disables thinking output. The validation fraction is 0.002, evaluation occurs every 20 steps, and the best checkpoint is loaded at the end. Inference uses deterministic decoding. An auxiliary sample-category prediction supplies the category used to retrieve empirical mean scores from routing training outcomes.

\subsection{Service Metadata and Eligibility}
Execution requirements are obtained from request processing and deployment metadata, including context length, attachment modalities, and required tool support. Eligibility checks compare these requirements with availability and configured service attributes. Optional limits on price rates and first-token latency use the available metadata. Missing values do not independently exclude a candidate. The generation-cost preference is applied separately as a soft selection preference.

\section{Model Profile Construction and Scoring Details}
\label{supp:measurement}

\subsection{Outcome Profiles and Missing-Cell Handling}
Outcome profiles aggregate observed task scores according to capability weights and profiler-assigned difficulty, as defined in the main paper. Both successes and failures contribute, and graded scores retain their original scale. Because one task can require several capabilities, its outcome may inform several cells. Each cell describes performance on tasks that demand the corresponding capability at a specified difficulty.

For an unobserved cell, lookup uses the nearest observed difficulty level and prefers the higher level in a tie. If a capability has no observed level, it contributes zero. The other capability weights retain their original values rather than being renormalized.

\subsection{Response Assessment Profiles}
\label{supp:response-assessment}
The assessment channel follows the general use of LLMs to evaluate response quality~\citep{zheng2023judging}. Its reference bank is sampled from the routing training partition and contains the available model responses for those requests. DeepSeek-V4-Flash annotates item requirements, and GPT-5.4 assesses responses against references and task rubrics. The requirement annotations share the capability schema of the deployed profiler but assign their weights independently. The capability-specific rubrics and aggregation procedure below define the response-assessment channel used by RouteAlign.

Let $w_{ik}$ denote the annotated requirement weight of item $i$ for capability $k$, and $z_{imk}\in[0,10]$ its assessment score for model $m$. The index set $\mathcal{J}_{m,k,\ell}$ contains valid assessments assigned to difficulty $\ell$ with positive requirement weight, evaluator relevance above 0.5, and a nonnegative response score. For nonempty support,
\begin{equation}
 A^{\mathrm{judge}}_{m,k,\ell}
 =\frac{\sum_{i\in\mathcal{J}_{m,k,\ell}}w_{ik}z_{imk}}
 {\sum_{i\in\mathcal{J}_{m,k,\ell}}w_{ik}}.
 \label{eq:judge-profile}
\end{equation}
The two measurement channels retain their native scales: $[0,1]$ for task outcomes and $[0,10]$ for response assessments. Separate learned coefficients combine them with the other evidence in the quality scorer. The common coordinates align both channels with task demand while preserving their distinct measurement scales.

\subsection{Retrieval and Feature Definitions}
\label{supp:features}
Retrieval uses sentence embeddings to compare tasks semantically~\citep{reimers2019sentence}. We use the frozen \texttt{all-MiniLM-L6-v2} checkpoint and normalize its 384-dimensional outputs. Each neighborhood contains $k_{\mathrm{nn}}=150$ training queries, with weights proportional to $\exp(-\beta\delta(x,x_j))$ for cosine distance $\delta$ and $\beta=5$. Weights sum to one over the neighborhood. The same retrieved records supply scalar quality, capability-conditioned quality, and generation-cost estimates. A capability estimate falls back to scalar neighborhood quality when its weighted support is at most $10^{-6}$.

Write $p_m=p_{\mathrm{nn}}(x,m)$ for scalar local quality, $b(x,m)$ for the training mean in the predicted sample category, and $g(x,m)$ for the mean in that category at difficulty $d(x)$. When the category-by-difficulty stratum is empty, $g(x,m)$ falls back to $b(x,m)$. Together with the compatibility scores defined in the main paper, these quantities form the ten-feature vector
\begin{equation}
 \begin{aligned}
 f(x,m)=\bigl[Q^{\mathrm{out}},Q^{\mathrm{judge}},g,b,p_m,
 Q^{\mathrm{local}},p_mD,\delta_m,\eta_m,|p_m-g|\bigr]^{\top}.
 \end{aligned}
 \label{eq:supp-features}
\end{equation}
Arguments $(x,m)$ are suppressed on the right-hand side. The margin $\delta_m$ subtracts the highest local estimate among other eligible candidates from $p_m$; the reference value is zero when only one candidate is eligible. The indicator $\eta_m$ records whether a candidate attains the highest local estimate. Selection ties follow a deterministic candidate order.

Before normalization, candidate separation is the maximum local quality minus its mean across eligible candidates. The benchmark divides this statistic by its mean over the corresponding feature-extraction batch to obtain $D(x)$, separately for training and evaluation inputs. All constituent quality estimates come from training-partition evidence. The normalization therefore depends on the requests included in the feature-extraction batch; the serving endpoint uses the request-level configuration in Section~\ref{supp:runtime}.

\subsection{Scorer Fitting and Missing Records}
Ridge regression uses penalty $\alpha=1$, native feature scales, and no intercept, with the recorded task score as the target for each training query-model pair. Retrieval features use leave-one-out neighborhoods. Global profiles and category priors use the complete training partition.

Absent records receive zero in dense retrieval scores, generation costs, and regression targets. Global outcome profiles aggregate observed records only, so a missing response contributes no observation to a profile cell. Feature ablations retain these accounting rules.

\section{Dataset Details and Baseline Implementations}
\label{supp:evaluation}

This section reports the evaluation split, response coverage, and baseline configurations used in our experiments. Citations identify the source methods; the implementation choices and fitting budgets specify the comparisons reported in the main paper.

\subsection{Train-Test Split}
Table~\ref{tab:data-split} reports the shared split obtained with seed 42 and a 70\% training fraction. All responses to a request remain in the same partition. Outcome profiles, response-assessment items, conditional means, retrieval references, and regression targets use training records. A policy selects a candidate before accessing the corresponding test outcome.

\begin{table}[t]
 \centering
 \captionsetup{width=0.48\linewidth}
 \caption{Training and test set sizes.}
 \label{tab:data-split}
 \small
 \setlength{\tabcolsep}{3pt}
 \begin{tabular*}{0.48\linewidth}{@{\extracolsep{\fill}}lrr@{}}
 \toprule
 Dataset & Train & Test \\
 \midrule
 AIME & 42 & 18 \\
 ArenaHard & 525 & 226 \\
 GPQA & 138 & 60 \\
 HLE & 1511 & 648 \\
 LiveCodeBench & 738 & 317 \\
 LiveMathBench & 84 & 37 \\
 MMLU-Pro & 2100 & 900 \\
 SimpleQA & 3028 & 1298 \\
 SWE-bench & 350 & 150 \\
 $\tau^2$-bench & 194 & 84 \\
 \midrule
 Total & 8,710 & 3,738 \\
 \bottomrule
 \end{tabular*}
\end{table}

\subsection{Candidate Coverage and Selection Counts}
Table~\ref{tab:model-coverage} records available test responses and RouteAlign selections in the fixed catalog. Sample-Avg averages scores over requests, whereas Dataset-Avg weights dataset means equally. The benchmark adapter rounds dataset means to two decimals. Both metrics retain graded task scores.

Direct-candidate summaries use available records and datasets. GPT-5 covers all 3,738 test requests, GPT-5-Chat has no $\tau^2$-bench records, and several other candidates have one or two missing records. Every RouteAlign selection has an observed response, producing a total task score of 2,357.5 and generation expenditure of \$110.4199 before rounding. AVG LLM in the main paper averages the 13 individual-candidate summaries. Random Router selects uniformly from the candidates for each request using seed 42.

\begin{table*}[t]
 \centering
 \captionsetup{width=0.82\linewidth}
 \caption{Candidate coverage and fixed-catalog selection counts. The selected candidates have observed outcomes for all 3,738 test requests.}
 \label{tab:model-coverage}
 \small
 \setlength{\tabcolsep}{5pt}
 \begin{tabular*}{0.82\linewidth}{@{\extracolsep{\fill}}lrrr@{}}
 \toprule
 Candidate checkpoint & Records & Datasets & Selections \\
 \midrule
 \texttt{intern-s1} & 3737 & 10 & 2 \\
 \texttt{deepseek-v3-0324} & 3737 & 10 & 12 \\
 \texttt{deepseek-v3.1-terminus} & 3737 & 10 & 37 \\
 \texttt{deepseek-r1-0528} & 3737 & 10 & 181 \\
 \texttt{qwen3-235b-a22b-2507} & 3737 & 10 & 807 \\
 \texttt{qwen3-235b-a22b-thinking-2507} & 3737 & 10 & 89 \\
 \texttt{kimi-k2-0905} & 3737 & 10 & 4 \\
 \texttt{glm-4.6} & 3737 & 10 & 14 \\
 \texttt{gpt-5-chat} & 3653 & 9 & 1 \\
 \texttt{gemini-2.5-flash} & 3737 & 10 & 2 \\
 \texttt{claude-sonnet-4} & 3736 & 10 & 137 \\
 \texttt{gemini-2.5-pro} & 3737 & 10 & 600 \\
 \texttt{gpt-5} & 3738 & 10 & 1852 \\
 \bottomrule
 \end{tabular*}
\end{table*}

\subsection{Preference-Based and Contrastive Routers}
RouteLLM~\citep{ong2025routellm} uses matrix factorization with frozen query embeddings and learned candidate vectors, fitted on 50,000 preference pairs. Our multi-candidate adaptation selects the highest candidate score; the original method applies a threshold to choose between a stronger and a weaker model. RouterDC~\citep{chen2024routerdc} uses mDeBERTa-v3-base, candidate embeddings, sample-model and sample-sample contrastive objectives, KMeans semantic groups, and 6,000 training examples; inference selects by cosine similarity. Neither evaluated inference rule applies a separate monetary penalty.

\subsection{Graph-Based and Cluster-Based Routers}
GraphRouter~\citep{feng2025graphrouter} represents tasks, queries, and candidates as nodes and uses two edge-aware graph convolution layers with masked-edge prediction. Its three configurations construct training targets from cost-oriented, balanced, and quality-oriented utility, respectively. AvengersPro~\citep{zhang2025avengerspro} summarizes quality and efficiency within query clusters. Our evaluated implementation uses balance parameter 0.5 and aggregates reciprocal candidate ranks from nearby clusters using exponential distance weights. This is the rank-based adaptation of cluster utility scoring used in the reported comparison.

\subsection{Classification-Based and Retrieval-Based Routers}
The classification baselines follow the KNNRouter, SVMRouter, and MLPRouter families provided by LLMRouter~\citep{feng2026llmrouter}. In our evaluation, they use frozen Longformer embeddings and labels identifying the highest-scoring training candidate. Their decision rules are neighbor-label voting, RBF-kernel classification, and a feed-forward classifier, respectively. Neighbor-label voting predicts a candidate identity; neighborhood performance estimation~\citep{li2025knnrouting} instead averages scores separately for each candidate. This distinction separates the KNNRouter comparison from the local quality estimates used by RouteAlign.

The three embedding-based retrieval policies use the \texttt{llm-semantic-router} checkpoints \path{multi-modal-embed-small}, \path{multi-modal-embed-large}, and \path{mmbert-embed-32k-2d-matryoshka}. MM-Embed-S and MM-Embed-L use 384- and 768-dimensional text embeddings, respectively. The mmBERT-32K configuration uses 768 dimensions, all 22 layers, and a 2,048-token input limit. Each policy averages candidate scores over cosine neighbors and breaks ties using lower neighbor cost. All comparison policies select before response generation under the same benchmark partition.

\section{Feature Ablations and Selection by Category Means}
\label{supp:additional-results}

\subsection{Feature Groups and Removal Rules}
The main paper includes every evaluated feature and cost-preference ablation. Feature ablations delete specified columns from Equation~\eqref{eq:supp-features} and refit ridge regression on the same training partition with leave-one-out retrieval features. The encoder and definitions of retained features remain fixed, and global profiles and category priors continue to use the full training partition. The grouped removals are:
\begin{itemize}
 \item All local quality evidence: $p_m$, $Q^{\mathrm{local}}$, $p_mD$, $\delta_m$, $\eta_m$, and $|p_m-g|$. The local generation-cost estimate remains available.
 \item Relative local evidence: $p_mD$, $\delta_m$, $\eta_m$, and $|p_m-g|$.
 \item Requirement-conditioned scores: $Q^{\mathrm{out}}$, $Q^{\mathrm{judge}}$, $g$, and $Q^{\mathrm{local}}$.
 \item Global profile compatibility: $Q^{\mathrm{out}}$ and $Q^{\mathrm{judge}}$.
\end{itemize}
Individual ablations of the scalar neighborhood estimate, sample-category prior, local capability conditioning, response-assessment compatibility, category-by-difficulty prior, and outcome compatibility delete $p_m$, $b$, $Q^{\mathrm{local}}$, $Q^{\mathrm{judge}}$, $g$, and $Q^{\mathrm{out}}$, respectively. Derived columns that remain in the feature vector keep their original definitions. In particular, deleting $p_m$ retains the four relative local features, and deleting $g$ retains $|p_m-g|$. These ablations measure the conditional contribution of the designated input columns.

Refitting redistributes predictive weight across retained estimates, so changes for overlapping removal groups are conditional on the remaining features. The cost-preference ablation retains the fitted quality estimator and sets $\lambda=0$.

\subsection{Category-Based Selection with and without Difficulty Conditioning}
These comparisons test whether empirical means for task categories are sufficient for selection. Using the same training evidence, the two reference policies select the candidate with the highest sample-category mean or category-by-difficulty mean, respectively, without the shared scorer. Table~\ref{tab:alignment-ablation} compares these rules with the complete RouteAlign policy.

\begin{table}[t]
 \centering
 \captionsetup{width=0.70\linewidth}
 \caption{Comparison with selection by category means. Each reference policy selects the candidate with the highest empirical mean score in the predicted category or category-and-difficulty group, without the shared scorer. Quality is in percent.}
 \label{tab:alignment-ablation}
 \footnotesize
 \setlength{\tabcolsep}{2pt}
 \begin{tabular*}{0.70\linewidth}{@{\extracolsep{\fill}}lrrr@{}}
 \toprule
 Selection rule & D-Avg. & S-Avg. & Cost (\$) \\
 \midrule
 \textbf{Full RouteAlign} & \textbf{68.64} & \textbf{63.07} & \textbf{110.42} \\
 Sample-category mean & 66.35 & 61.49 & 126.50 \\
 Category and difficulty mean & 66.42 & 61.45 & 110.68 \\
 \bottomrule
 \end{tabular*}
\end{table}

Conditioning on both category and difficulty yields 66.42\% Dataset-Avg and 61.45\% Sample-Avg at \$110.68 in generation charges. RouteAlign attains gains of 2.22 and 1.62 percentage points, respectively, at \$110.42. Selection by category alone yields 66.35\% Dataset-Avg and 61.49\% Sample-Avg at \$126.50. The complete policy therefore achieves higher quality than either rule based on category means, including the category-by-difficulty rule at similar observed generation cost.

These results support integrating complementary evidence beyond category means. The main-paper ablations examine how individual feature groups contribute within the fitted combination.

\section{Runtime Configurations and Measurements}
\label{supp:runtime}

\subsection{Selection with Cached Requirements and Embeddings}
The runtime table in the main paper measures selection over 3,738 test requests. RouteAlign reuses cached requirement profiles and embeddings; the measured operations are retrieval, feature construction, and candidate scoring. Offline preparation, requirement prediction, embedding construction, and candidate answer generation are excluded. The reported generation charges cover candidate responses and exclude the execution cost of this selection stage.

\subsection{Endpoint Latency with Online Requirement Prediction}
The separate endpoint runs the 0.6B requirement profiler and selects candidates using a 500-request reference bank. It computes candidate separation independently for each request as the unnormalized maximum local quality minus the mean. When sample-category metadata is absent, difficulty-level and global reference means replace the category-by-difficulty and category means, respectively. These settings define the request-level serving configuration.

The reported run processes 96 requests at concurrency 32. Timing covers requirement prediction and selection, ending before candidate answer generation. The resulting median latency is 0.60 seconds, the 95th-percentile latency is 0.74 seconds, and throughput is 51.2 requests per second, with no observed errors or fallbacks.

\section{Scorer Reuse under Catalog Changes}
\label{supp:catalog-visualization}
\begin{figure*}[t]
 \centering
 \includegraphics[width=\textwidth]{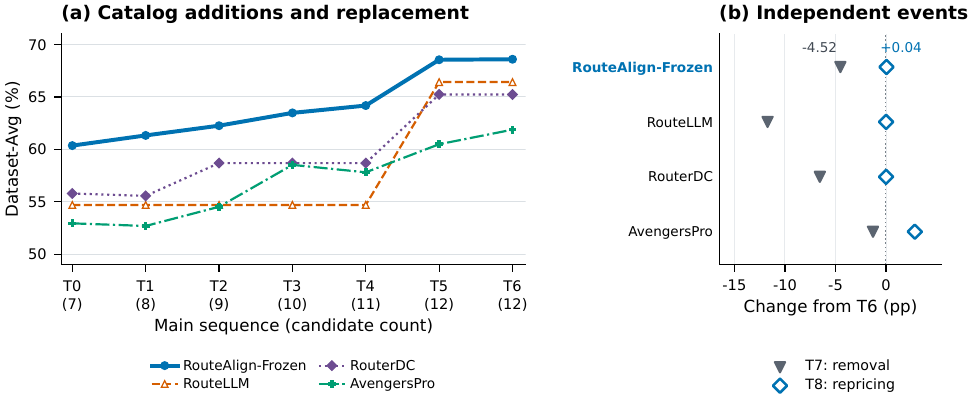}
 \caption{Selection quality under controlled catalog updates. (a) Dataset-Avg along T0 through T6, with candidate counts in parentheses. (b) Changes in percentage points for the independent T7 removal and T8 repricing branches relative to T6. RouteAlign-Frozen retains the scorer fitted on the seven T0 candidates and refreshes candidate evidence after each update.}
 \label{fig:pool-evolution}
\end{figure*}

Figure~\ref{fig:pool-evolution} presents the catalog measurements reported in the main paper. Panel (a) follows sequential additions and replacement; panel (b) separates removal and repricing into independent branches. The figure complements the update protocol and result table by showing how each event changes selection quality under the specified adaptation procedures.

Along T0 through T6, RouteAlign-Frozen retains the coefficients fitted on the seven T0 candidates. Newly admitted candidates are evaluated on training-partition reference tasks to obtain their measurements and profiles. Dataset-Avg increases from 60.36\% at T0 to 68.55\% after GPT-5 is added at T5, and to 68.60\% after the replacement at T6. These changes show that candidate measurements can be incorporated within the existing requirement coordinates and feature definitions while retaining the learned comparison rule.

The two events in panel (b) are measured against T6. Removing GPT-5 at T7 lowers RouteAlign-Frozen by 4.52 percentage points to 64.08\%, quantifying the effect of withdrawing this candidate. Halving the price of Gemini 2.5 Flash at T8 raises Dataset-Avg by 0.04 points to 68.64\%, with GPT-5 retained in the catalog.

RouteLLM and RouterDC retain checkpoints trained on all 13 candidates and mask unavailable outputs. RouteAlign-Frozen is fitted only on T0, while AvengersPro recomputes clusters and rankings for the current catalog. New candidate measurements come from reference evaluations performed before those candidates enter selection. All RouteAlign-Frozen results, including T8, use the T0-fitted coefficients; the fixed-catalog experiment uses coefficients fitted on all 13 candidates.

\section{Decision-Model Implementations and Detailed Results}
\label{supp:jev}

\subsection{Backend Implementations}
The comparison uses local adaptations of the five open projects in Table~\ref{tab:jev-backends}, instantiated at nine model sizes. Kev uses LoRA adapters and a pointer head, Open-Jev uses LoRA with a scalar decision head, and Decider uses fully fine-tuned models with option-letter logit readout. SemIf and OpenJev-mjdileep read pretrained-model logits. The experiments use the Qwen3.5-based versions listed below. All nine backends are evaluated without additional fine-tuning on the routing workload.

\begin{table*}[t]
 \centering
 \caption{Open implementations used in the Jev-style comparison. Project names link to their repositories, and checkpoint identifiers specify the evaluated Qwen3.5-based models. SemIf and OpenJev-mjdileep use pretrained base checkpoints directly.}
 \label{tab:jev-backends}
 \footnotesize
 \setlength{\tabcolsep}{3pt}
 \begin{tabular*}{\textwidth}{@{\extracolsep{\fill}}llp{0.23\textwidth}p{0.39\textwidth}@{}}
 \toprule
 Project & Sizes & Decision mechanism & Checkpoint identifiers \\
 \midrule
 \href{https://github.com/jaredpalmer/kev}{Kev} & 0.8B, 4B, 9B & LoRA and pointer head & \path{jaredpalmer/kev-0.8b}, \path{jaredpalmer/kev-4b}, \path{jaredpalmer/kev-9b} \\
 \href{https://github.com/Zefan-Cai/Open-Jev}{Open-Jev} & 2B, 9B & LoRA and scalar head & \path{ZefanCai/Open-Jev-2B}, \path{ZefanCai/Open-Jev-9B} \\
 \href{https://github.com/Mapika/decider}{Decider} & 2B, 4B & Full fine-tuning and letter logits & \path{Mapika/decider-2b}, \path{Mapika/decider-4b} \\
 \href{https://github.com/TheoLeeCJ/SemIf-OpenJev}{SemIf} & 4B & Pretrained letter logits & \path{Qwen/Qwen3.5-4B} \\
 \href{https://github.com/mjdileep/OpenJev}{OpenJev-mjdileep} & 0.8B & Pretrained affirmative-answer logits & \path{Qwen/Qwen3.5-0.8B} \\
 \bottomrule
 \end{tabular*}
\end{table*}

The local adapter follows the state-and-questions structure of the \href{https://docs.typesafe.ai/introduction}{Jev decision interface}. Open-Jev uses a locally constructed candidate prompt following the published scoring interface; Decider follows its model-card protocol. Kev processes question branches as separate causal sequences in a batch to accommodate the recurrent components of its backbone. Its state and combined state-question limits are 384 and 1,024 tokens, respectively. Released temperature parameters are loaded with the checkpoints.

\subsection{Requirement Prediction and Candidate Scoring}
One decision call contains 12 questions: ten four-level capability-demand scores, one L1--L4 difficulty choice, and one tool-requirement judgment. Each capability score is its expected ordinal level divided by three. Values below 0.05 are removed before the remaining weights are normalized. Tool use is required when its affirmative probability exceeds 0.5. The adapter then applies metadata filtering and matches candidates by the primary capability and predicted difficulty, using its configured tier fallback.

Candidate scoring receives the query truncated to 1,200 characters and a task summary. Each candidate description contains up to five capabilities selected by current demand, their difficulty-specific measurements, specialty tags, and service attributes, within a 700-character profile budget. The decision score maps the expected four-level performance rating to $[0,10]$. For decision score $J(x,m)$ and explicit profile compatibility $C(x,m)$, the final score is
\begin{equation}
 H_\rho(x,m)=\rho J(x,m)+(1-\rho)C(x,m),
 \qquad \rho\in\{0,0.5,1\}.
 \label{eq:jev-mixture}
\end{equation}
Here $C(x,m)$ sums the adapter's profile measurements weighted by the backend-predicted requirements at the predicted difficulty. At $\rho=0$, compatibility supplies the candidate score, while the backend determines the task requirements and resulting candidate set. The mixtures combine backend predictions with supplied global profile measurements; RouteAlign additionally uses retrieved local outcomes and its fitted ten-feature scorer. The shared workload and catalog support comparison of these complete selection policies.

The experiment saves both component scores and replays the three mixtures over the same candidate sets. Outcomes and generation charges are looked up only after selection. Every backend covers all 3,738 test requests, including successfully retried records. The implementation uses PyTorch 2.9.1, Transformers 5.3.0, PEFT 0.20.0, and bfloat16 inference on one A800 GPU.

\subsection{Complete Mixture Results}
Table~\ref{tab:jev-mixtures} provides quality and generation expenditure for all 27 configurations. The best configuration under both aggregate metrics is Decider-4B with explicit compatibility scoring. RouteAlign improves on it by 6.44 Dataset-Avg points and 5.22 Sample-Avg points at 9.7\% lower generation expenditure. Intermediate mixtures help some backends, including Open-Jev-2B and SemIf-4B, while Decider-4B performs best at $\rho=0$. RouteAlign's aggregate advantage is sustained across this variation in backend and scoring mixture.

\begin{table*}[t]
 \centering
 \caption{All scoring mixtures for the open Jev-style backends. D and S denote Dataset-Avg and Sample-Avg (\%); cost is in US dollars. The last column is mean requirement-prediction and scoring time on one A800, shared by the three offline replays.}
 \label{tab:jev-mixtures}
 \footnotesize
 \setlength{\tabcolsep}{2.5pt}
 \resizebox{\textwidth}{!}{%
 \begin{tabular}{@{}l*{10}{r}@{}}
 \toprule
 & \multicolumn{3}{c}{$\rho=0$} & \multicolumn{3}{c}{$\rho=0.5$} & \multicolumn{3}{c}{$\rho=1$} & \\
 \cmidrule(lr){2-4}\cmidrule(lr){5-7}\cmidrule(lr){8-10}
 Backend & D & S & Cost & D & S & Cost & D & S & Cost & Time (ms) \\
 \midrule
 Kev-0.8B & 48.08 & 47.14 & 36.58 & 51.39 & 49.02 & 38.11 & 51.83 & 49.65 & 39.33 & 1268.9 \\
 OpenJev-mjdileep-0.8B & 48.76 & 45.77 & 48.61 & 49.86 & 45.80 & 51.56 & 47.44 & 45.85 & 52.57 & 969.7 \\
 Open-Jev-2B & 47.70 & 46.54 & 50.21 & 51.78 & 49.83 & 49.90 & 48.93 & 48.86 & 39.52 & 1544.0 \\
 Decider-2B & 56.42 & 46.63 & 123.39 & 56.11 & 47.27 & 108.88 & 55.00 & 47.11 & 95.63 & 593.4 \\
 SemIf-4B & 49.35 & 51.47 & 30.99 & 54.29 & 52.62 & 33.55 & 53.43 & 51.82 & 29.32 & 1727.8 \\
 Kev-4B & 60.78 & 54.70 & 105.57 & 61.85 & 55.35 & 105.63 & 61.43 & 55.43 & 102.97 & 1665.8 \\
 Decider-4B & 62.20 & 57.85 & 122.28 & 61.23 & 57.45 & 113.84 & 60.61 & 57.08 & 109.32 & 1727.5 \\
 Open-Jev-9B & 48.78 & 45.65 & 79.49 & 50.20 & 46.11 & 65.00 & 53.44 & 47.61 & 44.66 & 3212.7 \\
 Kev-9B & 59.48 & 55.93 & 121.22 & 61.00 & 56.45 & 110.96 & 60.50 & 56.51 & 111.32 & 1713.4 \\
 \bottomrule
 \end{tabular}%
 }
\end{table*}

\subsection{Performance across Task Groups}
Table~\ref{tab:jev-per-dataset} disaggregates direct decision scoring. RouteAlign achieves the highest score on seven of the ten datasets in this comparison, including information-intensive, code-generation, and interactive workloads. Decider is stronger on AIME, Kev-9B on GPQA, and Kev-4B on SWE-bench. The aggregate advantage reflects gains across several task groups, complementing the comparison across scoring mixtures in Table~\ref{tab:jev-mixtures}.

\begin{table*}[t]
 \centering
 \caption{Per-dataset quality (\%) with direct decision scoring ($\rho=1$). Bold marks the highest score in each column among these configurations and RouteAlign.}
 \label{tab:jev-per-dataset}
 \footnotesize
 \setlength{\tabcolsep}{2.5pt}
 \resizebox{\textwidth}{!}{%
 \begin{tabular}{@{}l*{10}{r}@{}}
 \toprule
 Backend or policy & AIME & AH & GPQA & HLE & LCB & LMB & MMP & SQA & SWE & $\tau^2$ \\
 \midrule
 Kev-0.8B & 61.11 & 69.03 & 73.33 & 8.80 & 59.94 & 54.05 & 82.33 & 43.76 & 20.67 & 45.24 \\
 OpenJev-mjdileep-0.8B & 55.56 & 69.03 & 66.67 & 12.35 & 56.15 & 62.16 & 82.67 & 34.05 & 16.67 & 19.05 \\
 Open-Jev-2B & 55.56 & 67.92 & 63.33 & 8.18 & 59.94 & 62.16 & 81.67 & 44.30 & 15.33 & 30.95 \\
 Decider-2B & \textbf{88.89} & 70.35 & 70.00 & 15.43 & 70.98 & 62.16 & 83.22 & 29.43 & 22.67 & 36.90 \\
 SemIf-4B & 66.67 & 65.04 & 78.33 & 9.26 & 58.36 & 59.46 & 83.33 & 50.46 & 8.67 & 54.76 \\
 Kev-4B & 83.33 & 73.89 & 81.67 & 13.89 & 77.29 & 64.86 & 83.22 & 49.08 & \textbf{34.67} & 52.38 \\
 Decider-4B & \textbf{88.89} & 68.36 & 80.00 & 17.13 & 82.33 & 64.86 & 84.22 & 53.24 & 20.67 & 46.43 \\
 Open-Jev-9B & 55.56 & 72.79 & 75.00 & 9.26 & 53.63 & 70.27 & 83.22 & 36.44 & 24.67 & 53.57 \\
 Kev-9B & 72.22 & 74.12 & \textbf{85.00} & 15.28 & 76.03 & 70.27 & 83.22 & 52.70 & 27.33 & 48.81 \\
 \midrule
 \textbf{RouteAlign} & 83.33 & \textbf{78.10} & 81.67 & \textbf{26.08} & \textbf{83.60} & \textbf{81.08} & \textbf{86.89} & \textbf{58.47} & 33.33 & \textbf{73.81} \\
 \bottomrule
 \end{tabular}%
 }
 \par\smallskip
 \begin{minipage}{\linewidth}
 \footnotesize AH: ArenaHard; LCB: LiveCodeBench; LMB: LiveMathBench; MMP: MMLU-Pro; SQA: SimpleQA; SWE: SWE-bench.
 \end{minipage}
\end{table*}

\subsection{Backend Runtime}
The final column of Table~\ref{tab:jev-mixtures} measures sequential request processing on one A800, covering requirement prediction and candidate scoring before answer generation. The observed means range from 593.4 to 3,212.7 milliseconds. Runtime is recorded for the underlying inference run and shared by the three offline replays of its saved component scores. Section~\ref{supp:runtime} reports RouteAlign's cached selection and concurrent serving-endpoint measurements, with the operations and serving configuration specified for each.

\end{document}